\documentclass[prd,aps,twocolumn,showpacs,superscriptaddress,nofootinbib]{revtex4-1}
\usepackage{amsfonts,amsmath,amssymb,bm,dcolumn,mathrsfs,multirow}
\usepackage{graphicx,graphics,latexsym,xspace,color,isotope}
\usepackage[colorlinks=true,linkcolor=blue,citecolor=green,urlcolor=blue]{hyperref}

\newcommand{\be}{\begin{equation}}
\newcommand{\ee}{\begin{equation}}
\newcommand{\bea}{\begin{eqnarray}}
\newcommand{\eea}{\end{eqnarray}}

\definecolor{red}{rgb}{0.8,0,0}
\definecolor{violet}{rgb}{0.4,0,0.4}
\definecolor{green}{rgb}{0,0.5,0.0}
\definecolor{navy}{rgb}{0.0,0.0,0.6}
\definecolor{orange}{rgb}{0.8,0.2,0.0}

\usepackage[normalem]{ulem}  

\usepackage[]{times}

\def\apj{ApJ}%
\def\apjs{ApJS}%
\def\aap{A\&A}%
\def\jcap{J. Cosmology Astropart. Phys.}%
\def\mnras{MNRAS}%
\def\na{New Astron.}%
\def\prc{Phys.~Rev.~C}%
\def\prd{Phys.~Rev.~D}%
\def\prl{Phys.~Rev.~Lett.}%
\def\jcap{JCAP}%
\def\nphysb{Nucl.~Phys.~B}%
\begin{document}

\title{Axion cooling of neutron stars. II. Beyond hadronic axions}

\author{Armen Sedrakian}
  \affiliation{Frankfurt Institute for 
  Advanced Studies, Ruth-Moufang str.\ 1, D-60438 Frankfurt-Main,
  Germany }

\begin{abstract}
  We study the axion cooling of neutron stars within the
  Dine-Fischler-Srednicki-Zhitnitsky (DFSZ) model, which allows for
  tree level coupling of electrons to the axion {and locks the
    Peccei-Quinn charges of fermions via an angle parameter}. This
  extends our previous study [Phys. Rev. D 93, 065044 (2016)] limited
  to hadronic models of axions. We explore the two-dimensional space
  of axion parameters within the DFSZ model by comparing the
  theoretical cooling models with the surface temperatures of a few
  stars with measured surface temperatures. It is found that axions
  masses $m_a\ge 0.06$ to 0.12 eV can be excluded by x-ray observations
  of thermal emission of neutron stars (in particular by those of Cas
  A), the precise limiting value depending on the angle parameter of
  the DFSZ model.  {It is also found that axion emission by
    electron bremsstrahlung in neutron star crusts is negligible
    except for the special case where neutron Peccei-Quinn charge is
    small enough, so that the coupling of neutrons to axions can be
    neglected. }
\end{abstract}

\date{\today}
\maketitle

\section{ Introduction}  

Axions were suggested four decades
ago~\cite{1978PhRvL..40..279W,1978PhRvL..40..223W} to solve the
strong-{\it CP} problem in QCD~\cite{1976PhRvL..37....8T}. They are one of
the viable candidates for the cold dark matter in cosmology and can
play an important role in the stellar astrophysics.  Axions are
identified with the pseudo-Goldstone bosons which emerge through the
spontaneous breaking of the approximate Peccei-Quinn (PQ) global
$U(1)_{PQ}$ symmetry~\cite{1977PhRvL..38.1440P,2008LNP...741....3P}.
Their coupling to the Standard Model (SM) particles is determined by a
decay constant $f_a$ and PQ charges of the SM particles.  For reviews
of searches of axions in experiments and limits on their properties
from astrophysics see Refs.~\cite{Ringwald2012,Giannotti2016,Irastorza2018}.

In a previous work~\cite{Sedrakian2016} (hereafter Paper I) the axion
cooling of neutron stars was studied on the basis of numerical
simulations, with the aim of placing constraints on the axion coupling
(or, equivalently, the mass $m_a$) through comparison of the
simulation results for neutron star surface temperatures with the
observed surface photon luminosities of a few well-studied objects. As
in the case of the Sun, solar-type stars, red giant stars, white
dwarfs, and supernovae constraints on axion properties can be obtained
by requiring that the coupling of axions to SM particles should not
alter significantly the agreement between theoretical models and
observations~\cite{2011PhRvD..84j3008R,Giannotti2017}. In Paper I the
PQ charges of constituents of neutron star matter were chosen
according to the {\it hadronic model} of axions, i.e., the
Kim-Shifman-Vainstein-Zakharov (KSVZ)
model~\cite{Kim1979,Shiftman1980}. In this model protons and neutrons
have nonzero PQ charges and, therefore, couple to the axion at the
tree level. On the contrary, electron's PQ charge is zero, i.e., the
axion does not couple to the electron at the tree level. In an
alternative Dine-Fischler-Srednicki-Zhitnitsky (DFSZ) axion
model~\cite{Dine1981,Zhitnitsky:1980tq} electrons have nonzero PQ
charges, therefore the electronic component of a neutron star can cool
by emitting axions. Furthermore, in the DFSZ model the couplings of
the SM particles depend only on an angle parameter and $f_a$, which
limits the parameter space of this model to a two-dimensional
plane. It is the purpose of this work to adapt and extend the
computations reported in Paper I to the DFSZ axion model. A new aspect
of this study is the additional axion emission through the electronic
component of the star, which contributes alongside the axion emission
by hadrons studied in detail in Paper I. { Another novelty is the
  ``locking'' of the fermionic PQ charges via an angle parameter in
  the DFSZ model, which restricts the parameter space and, therefore,
  facilitates the parameter study of cooling curves.}

{A number of complementary studies of axion cooling of neutron
  stars have recently used the data on compact central objects (CCOs)
  to place limits on the axion properties. The transient behavior of
  the Cas A has been studied in
  Refs.~\cite{Leinson2014JCAP,Hamaguchi2018} to this end and useful
  limits were obtained assuming that the data reflect {\it per se} the
  fast cooling of this object.  The cooling behavior of peculiarly
  ``hot'' CCO HESS J1731-347 has been analyzed in the context of
  axionic cooling in Ref.~\cite{Beznogov2018}. }

This paper is structured as follows: In Sec.~\ref{sec:microphysics} we
start with a brief review of axion emission processes, discuss the
axion coupling to SM particles within the DFSZ model and concentrate on the rate of
axion emission by electron bremsstrahlung in neutron star crusts.
Sec.~\ref{sec:simulations} discusses the simulation setup and the
resulting cooling tracks for a large array of models of neutron
stars. Our conclusions and an outlook are given in
Sec.~\ref{sec:conclusions}.  The natural units with $\hbar =c=k_B= 1$,
$\alpha = 1/137$ will be used unless stated otherwise.

\section{Microphysics of axion emission in neutron stars}
\label{sec:microphysics}

\subsection{Overview}

The focus of this work, from the microscopic point of view, is the
bremsstrahlung of axions by electrons which are scattered on nuclei in
neutron star crusts. This process has been initially studied in
Ref.~\cite{Iwamoto1984}. Improved rates which include many-body
correlations were derived later in
Refs.~\cite{Nakagawa1987,Nakagawa1988}. However, these rates have not
been implemented in cooling simulations of neutron stars previously;
the pioneering simulations of Umeda {\it et al.}~\cite{Umeda1998} contain
results obtained within the DFSZ model, but the electron
bremsstrahlung process has not been mentioned. As indicated above, our
simulations in Paper I were limited to hadronic KSVZ model which does
not couple the axions to electrons at the tree level. Nevertheless,
the electron bremsstrahlung of axions was considered in detail in the
context of cooling of white
dwarfs~\cite{1994APh.....2..175A,1995PhRvD..51.1495R,2001NewA....6..197C,2014JCAP...10..069M}
and appropriate limits on the electron-axion coupling were derived
from comparisons of white-dwarf cooling models and their
observations. We will discuss the implementation of this process in
the following subsection.

A leading axion emission process from the interiors of neutron stars
is the axion ($a$) bremsstrahlung by nucleons ($N$): $N+N\to N+N+a$.
It was studied in the context of type-II supernovae and the bounds on
axion properties were derived by requiring consistency between the
explosion energetics as well as energies of neutrinos observed in the
1987A event and energy drained by axion
emission~\cite{1988PhRvD..38.2338B,1989PhRvD..39.1020B,1990PhRvD..42.3297B,1996PhRvL..76.2621J,2001PhLB..499....9H}.
More recent work concluded that future supernova observations could
probe axion mass range $m_a \le 10^{-2}$ eV
\cite{2016PhRvD..94h5012F}. Axions may not free stream in supernovae
if their coupling to matter is large enough.
Burrows {\it et al.}~\cite{1990PhRvD..42.3297B} find that axions are trapped within a
newborn neutron star if the axion mass is larger than $10^{−2}$~eV.
This implies the existence of an ``axion sphere,'' i.e., a surface of
last interaction of axions with the ambient matter at the initial
stage of neutron star evolution. However the physics at the early
moments of neutron star cooling does not affect the following stages
of thermal evolution significantly, therefore our simulations are
started at a temperature at which axions and neutrinos are untrapped,
which is typically $T\simeq 5$~MeV.

Axion bremsstrahlung via Cooper pair-breaking-formation (PBF)
processes sets in after the nucleons undergo a superfluid phase
transition~\cite{2013NuPhA.897...62K,Sedrakian2016}. These have been
the dominant axion emission processes in the KSVZ model. Our previous
limits reflect the efficacy of these processes in cooling neutron
stars below the observed temperatures in the {\it neutrino cooling
  era}, which corresponds to the time span $0.1\le t\le 100$~kyr. It
is understood that their neutrino counterpart PBF
processes~\cite{1999A&A...343..650Y,2006PhLB..638..114L,2007PhRvC..76e5805S,2008PhRvC..77f5808K,2012PhRvC..86b5803S}
are sufficient to cool the stars towards their current observational
values.  {Interestingly, PBF processes can trap axions at the late
  stages of cooling if $f_{a} \leq 10^6$ GeV due to the inverse proton
  PBF, as has been pointed out in Ref.~\cite{Hamaguchi2018}. The
  $f_{a}$-values discussed below are all above this limit, therefore
  we will ignore the possibility of axions being trapped.  }

\subsection{DFSZ model of axion coupling to SM particles}
\label{sec:axion_SM}

The Lagrangian of axion field $a$ has the form
\bea \mathscr{L}_a = -\frac{1}{2}\partial_{\mu}
a\partial^{\mu} a + \mathscr{L}^{(N)}_{int}(\partial_{\mu} a,\psi_{N})
+ \mathscr{L}^{(L)}_{int}(a,\psi_L),
\eea
where the second and third terms describe the coupling of the axion
to the nucleonic ($\psi_N$) and leptonic fields ($\psi_L$) of the SM.
The second term is given explicitly by the interaction Lagrangian
\bea 
\mathscr{L }^{(N)}_{int} = \frac{1}{f_a} N^{\mu} \partial_{\mu} a,
\quad N^{\mu} =
\sum_{N} \frac{C_N}{2}  \bar\psi_N\gamma^{\mu}\gamma_5\psi_N,\quad \quad 
\eea
where $N\in n,p$ stands for neutron or proton, $N^{\mu}$ is the
baryon current, $f_a$ is the axion decay constant, and $C_N$ is the
PQ charge of baryon $N$.  The coupling of axions to electrons can be
written in the pseudoscalar form
\bea
\mathscr{L}^{(e)}_{int}(a,\psi_e)
= \frac{C_e}{2f_a}\bar\psi_e\gamma^{\mu}\gamma_5 \psi_e
(\partial_\mu a)
= -ig_{ae} \bar\psi_e\gamma_5\psi_e a, 
\eea 
where the Yukawa coupling is given by $g_{ae}= C_em_e/f_a$ with $m_e$
being the electron mass.  We will also use a ``fine-structure
constant'' associated with this coupling, which is defined as $\alpha_{ae} =g_{ae}^2/4\pi$.

The PQ charges for the proton and neutron are given by
generalized Goldberger-Treiman relations 
\bea C_p & =& (C_u-\eta)
\Delta_u+(C_d-\eta z) \Delta_d +(C_s-\eta w)\Delta_s,\\
C_n & =& (C_u-\eta) \Delta_d+(C_d-\eta z) \Delta_u 
+(C_s-\eta w)\Delta_s, 
\eea 
where $\eta = (1+z+w)^{-1}$, with $z = m_u/m_d$, $w = m_u/m_s$,
$\Delta_u =0.84\pm 0.02$, $\Delta_d = -0.43\pm 0.02$ and
$\Delta_s = -0.09\pm 0.02$. The main uncertainty arises from the
quark-mass ratios: $ 0.35\le z\le 0.6$ and $17\le m_s/m_d\le 22$. We
adopt below the following mean values: $z=0.5$ and $w= 0.025$.

In the DFSZ model, the PQ charges are given by 
\bea
C_e =C_d =C_s =\frac{\cos^2\beta}{3},\quad 
C_u = \frac{\sin^2\beta}{3}, 
\eea
where the angle $\beta$ is a free parameter. 
\begin{table}
\begin{tabular}{c  l  r   r   r}
\hline
\\
 & $\cos ^2\beta$ & $C_n$& $C_p$ & $C_e$ \\
\\
\hline
\\
 & 0.0   &  $-0.14$  &  $-0.13$  &   0.00  \\
   &0.25 &  $-0.04$  &  $-0.24$ &   0.08\\ 
   &0.5 &     0.06   &  $-0.36$ &   0.17  \\
   & 0.75 & 0.16  &   $-0.47$  &   0.25  \\
   & 1.0   & 0.26  &   $-0.58$ &   0.33  \\
\\
\hline 
\end{tabular}
\caption{The values of the axion-nucleon and  axion-electron coupling 
constants for various values of parameter  $\cos^2\beta$. 
}\label{tab:1}
\end{table}
\begin{figure}[tbh]
\begin{center}
\includegraphics[width=0.49\textwidth]{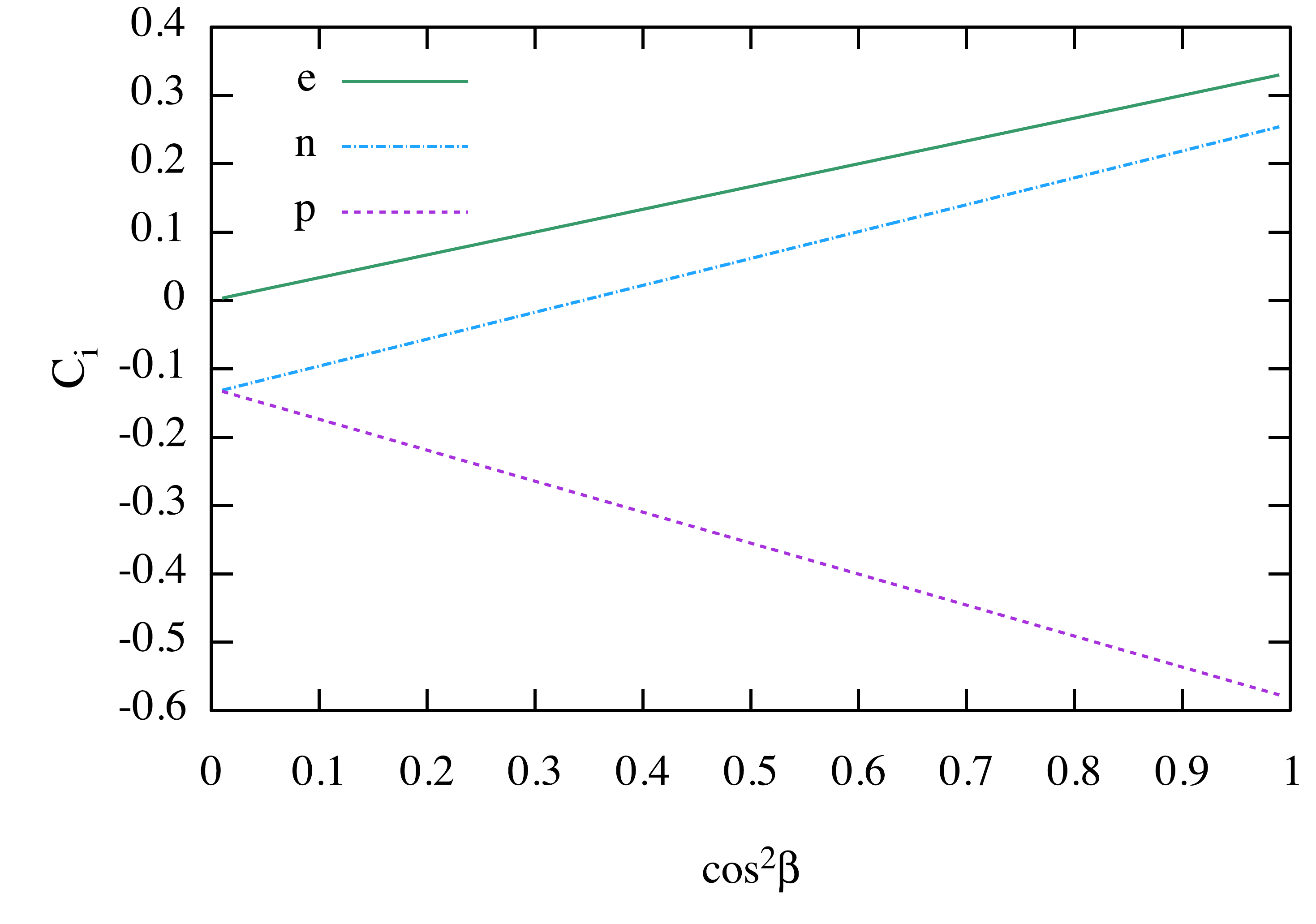}
\caption{ 
Dependence of the PQ charges of the electron ($e$),
neutron $(n)$ and proton ($p$) on the parameter $\cos^2\beta$. 
}
\label{fig:couplings}
\end{center}
\end{figure}

Finally, the axion mass is  given by 
\bea 
\label{eq:axion_mass}
 m_a = \frac{z^{1/2}}{1+z} \frac{f_\pi m_\pi}{f_a} 
= \frac{0.6~\textrm{eV}}{f_{a7}} ,
\eea 
where $f_{a7}= f_a/(10^{7}$\, GeV), the pion mass $m_\pi=135$ MeV, its
decay constant is$f_\pi = 92$ MeV, and $z=0.5$ as above. Note that
Eq.~(\ref{eq:axion_mass}) translates a lower bound on $f_a$ into an
upper bound on the axion mass. Table~\ref{tab:1} displays the set of
axion-fermion couplings for five values of the parameter $\cos
^2\beta$ which 
are used below to cover the relevant range of cooling 
simulations. In addition, we show in Fig.~\ref{fig:couplings} the same 
dependence in the full range $0\le \cos
^2\beta\le 1$.

\subsection{Axion bremsstrahlung emission in the crust}

At temperatures relevant for {\it neutrino cooling era} the dominant
cooling process associated with the electron component of the star is
the electron bremsstrahlung of neutrino--anti-neutrino pairs or axions
when electrons are scattered off the nuclei. For all relevant temperatures
and densities, ions are fully ionized and electrons form an ultrarelativistic, 
weakly interacting gas.  The correlations in the ionic component 
are characterized by the Coulomb
plasma parameter 
\bea\label{eq:Gamma}
\Gamma=\frac{e^2 Z^2}{Ta_i}\simeq 22.73
\frac{Z^2}{T_6}\bigg(\frac{\rho_6}{A}\bigg)^{1/3},
\eea
where $e$ is the elementary charge, $A$ and $Z$ are the mass number and
charge of a nucleus, $T$ is the temperature, $a_i=(4\pi n_i/3)^{-1/3}$
is the radius of the spherical volume per ion, $n_i$ the number
density of nuclei, $T_6$ is the temperature in units $10^6$ K, and
$\rho_6$ is the density in units of $10^6$ g cm$^{-3}$. The ionic
component is in the liquid state for values of
$\Gamma\leq\Gamma_m\simeq 180$. Otherwise, it forms a lattice, i.e.,
for $\Gamma>\Gamma_m$ and the electrons are scattering on the lattice
and phonons.  For a recent compilation of the
phase diagram of matter in the crust of a neutron star and its
dependence on the composition of matter see
Ref.~\cite{Harutyunyan2016}.

The axion emissivity can be written in the solid ($S$) and liquid
($L$) phases in the generic form~\cite{Iwamoto1984,Nakagawa1987,Nakagawa1988}
\bea
\label{eq:emissivity_1}
\epsilon_{L/S} &=& 
\frac{4\pi^2}{15} \frac{(\alpha Z )^2}{A}
\frac{\alpha_{ae} n_B }{\hbar^2c} 
\frac{(k_BT)^4}{(2c p_F)^2} \left(\frac{p_F}{m_e}\right)^2 \,  F_{L/S},
\eea
where for the sake of clarity we recovered the fundamental constants,
$p_F$ is the Fermi momentum of electrons, $n_B = A n_i$ is the nucleon
number density, and $F_{L/S}$ are correlation functions defined in
Refs.~\cite{Nakagawa1987,Nakagawa1988}. They depend (among other
factors) on the static structure factor of ions and the nuclear
form-factor of the nucleus. After substituting the numerical constants
one finds~\cite{Nakagawa1987,Nakagawa1988}
\bea
\label{eq:emissivity_brems}
\epsilon_{L/S}
&=& 
 1.08\,\rho\, \alpha_{ae,26} \frac{Z^2}{A} T_8^4\, 
\,F_{L/S}\, \textrm{[erg cm$^{-3}$ s$^{-1}$] },
\eea
where $\rho$ is the mass density, $T_{8} = T/(10^8 \, K)$ and
$\alpha_{ae26} = 10^{26}\alpha_{ae}$ with
$g_{ae} = \sqrt{4\pi\alpha_{ae}} = C_em_e/f_{a}\simeq 1.67\times
10^{-11}\cos^2\beta /f_{a7}$.
The correlation functions in the solid and liquid phases were obtained
through fits to the data provided in Fig.~3 of Ref.~\cite{Nakagawa1987}.
In the solid phase the contribution of the lattice is taken into
account, but the small phonon contribution is neglected, see Fig.~3 of
Ref.~\cite{Nakagawa1988}. For practical purposes, we use fits to these
computations which are given in the Appendix.

\section{Cooling simulations}
\label{sec:simulations}

To make our presentation self-contained we remind here the basic
assumptions underlying the strategy adopted in Paper I: (a) the
simulations are based on a conservative model of cooling of neutron
stars, which requires that the stellar models describing the data are
not massive enough to allow for fast cooling processes to occur.  This
requirement is based on the observation that fast cooling agents
appear only above certain density threshold which can be reached only
in massive compact stars. The light- to medium-mass neutron stars
within the mass range $1\le M/M_{\odot}\le 1.8$ are good candidates
for such cooling.  (b) The simulations are compared to observational
data for sources with estimated magnetic fields of the order of
canonical pulsar fields $B\simeq 10^{12}$~G and below. This ensures
that internal heating by strong magnetic
fields~\cite{2013MNRAS.434..123V} can be excluded.  (c) We continue to
use the {\sc NSCool} code~\footnote{The {\sc NSCool} code is available at:
  http://www.astroscu.unam.mx\\/neutrones/NSCool/.} with 
 its specific microphysics input to guarantee the easy reproduction of
our results and to benchmark the axion cooling of neutron stars (see
Paper I for details). The code has been extended to include all the
relevant axionic emission processes by hadrons and electrons as
discussed above.
\begin{figure*}
\begin{center}
\includegraphics[width=0.8\textwidth]{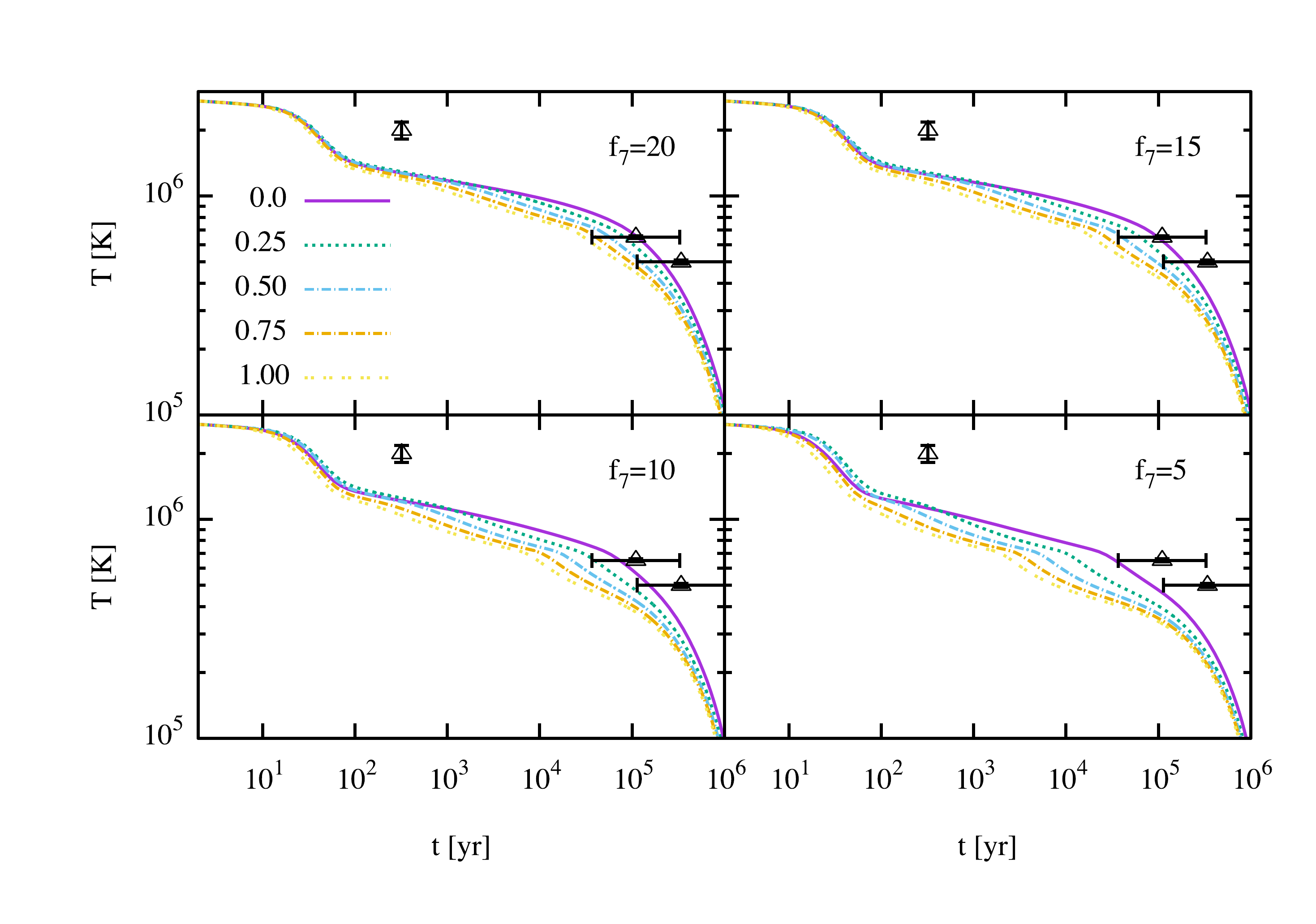}
\caption{ Cooling tracks of neutron star models with mass
  $M= 1.4M_{\odot}$ for the case of a nonaccreted iron envelope
  ($\eta = 0$). The data shows the surface temperatures inferred from
  the black-body fits to the x-ray emission of CCO in Cas A, PSR
  B0656+14 and Geminga. Each panel corresponds to fixed value of
  $f_{a7}$ as indicated. The values of PQ charges are specified in
  terms of $\cos^2\beta$ parameter, see Table~\ref{tab:1}.
}
\label{fig:2}
\end{center}
\end{figure*}
\begin{figure*}
\begin{center}
\includegraphics[width=0.8\textwidth]{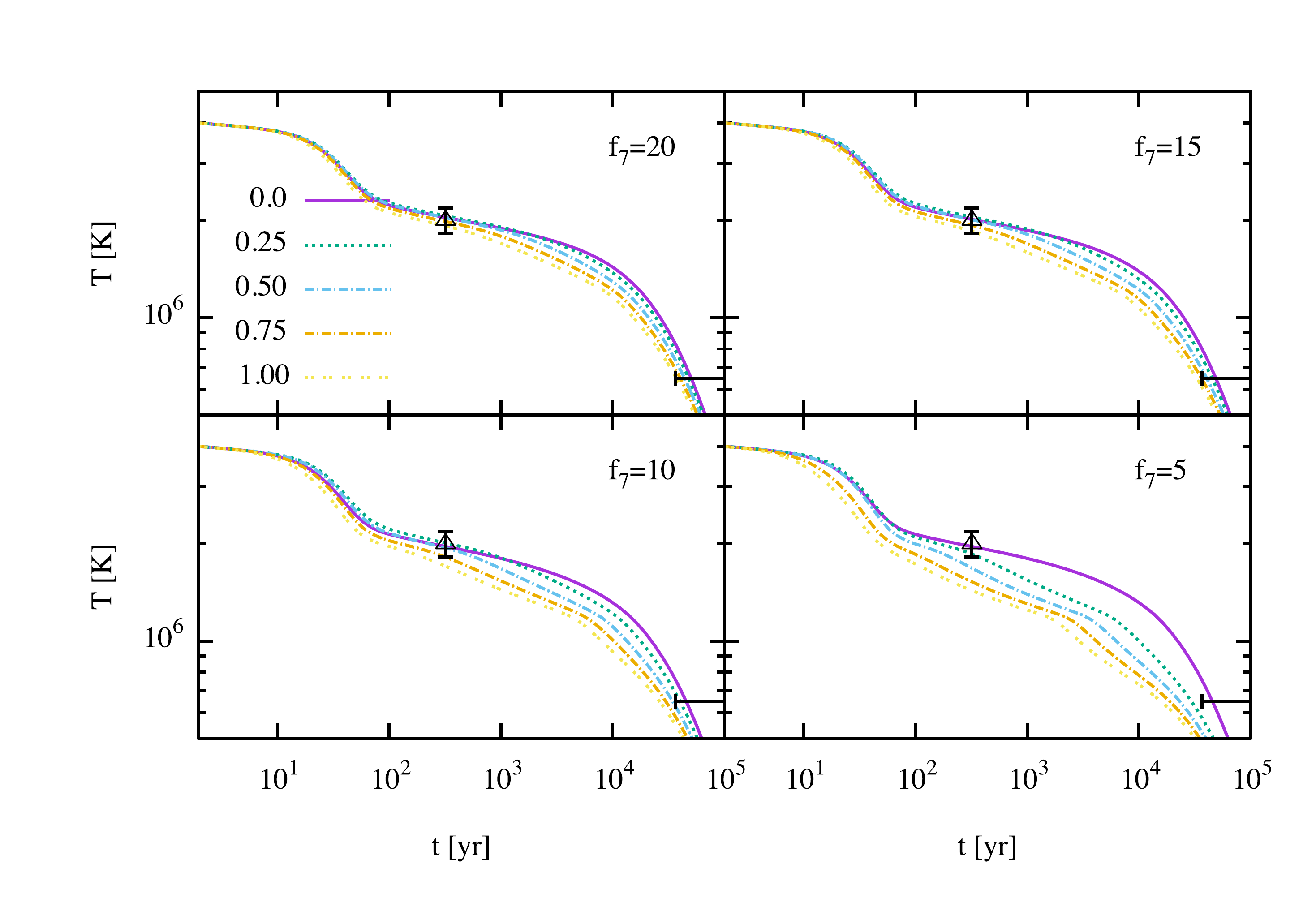}
\caption{Same as in Fig.~\ref{fig:2} but for $\eta =1$.}
\label{fig:3}
\end{center}
\end{figure*}

\subsection{Physics input and observational data}

The cooling code solves the energy balance and transport equations in
spherical symmetry, i.e., rotation and magnetic fields are
excluded. We use a generic relationship between the surface
temperature $ T_s$ and the temperature of the shell at density
$\rho_b= 10^{10}$g cm$^{-3}$ to avoid the problem of radiative
transport in the thin blanket lying below this density. This relation
is given by $ T_s^4 = g_sh(T)$, where $g_s$ is the surface gravity,
and $h$ is some function which depends on the temperature $T$, the
opacity of the blanket, and its equation of state. The surface
composition of a neutron star is modeled by the parameter $\eta$,
with $\eta = 0$ corresponding to a purely iron surface and $\eta\to 1$
to a light-element surface. Further details of the input physics can
be found in Ref.~\cite{2009ApJ...707.1131P} and in Paper I.
Throughout a cooling simulation, we extract the neutrino and axion
luminosities of our models, as well as the photon luminosity which is
given by the Stefan-Boltzmann law $L_{\gamma} = 4\pi \sigma R^2T_s^4$,
where $\sigma$ is the Stefan-Boltzmann constant, and $R$ is the radius
of the star.

\begin{figure*} \begin{center}
    \includegraphics[width=0.80\textwidth]{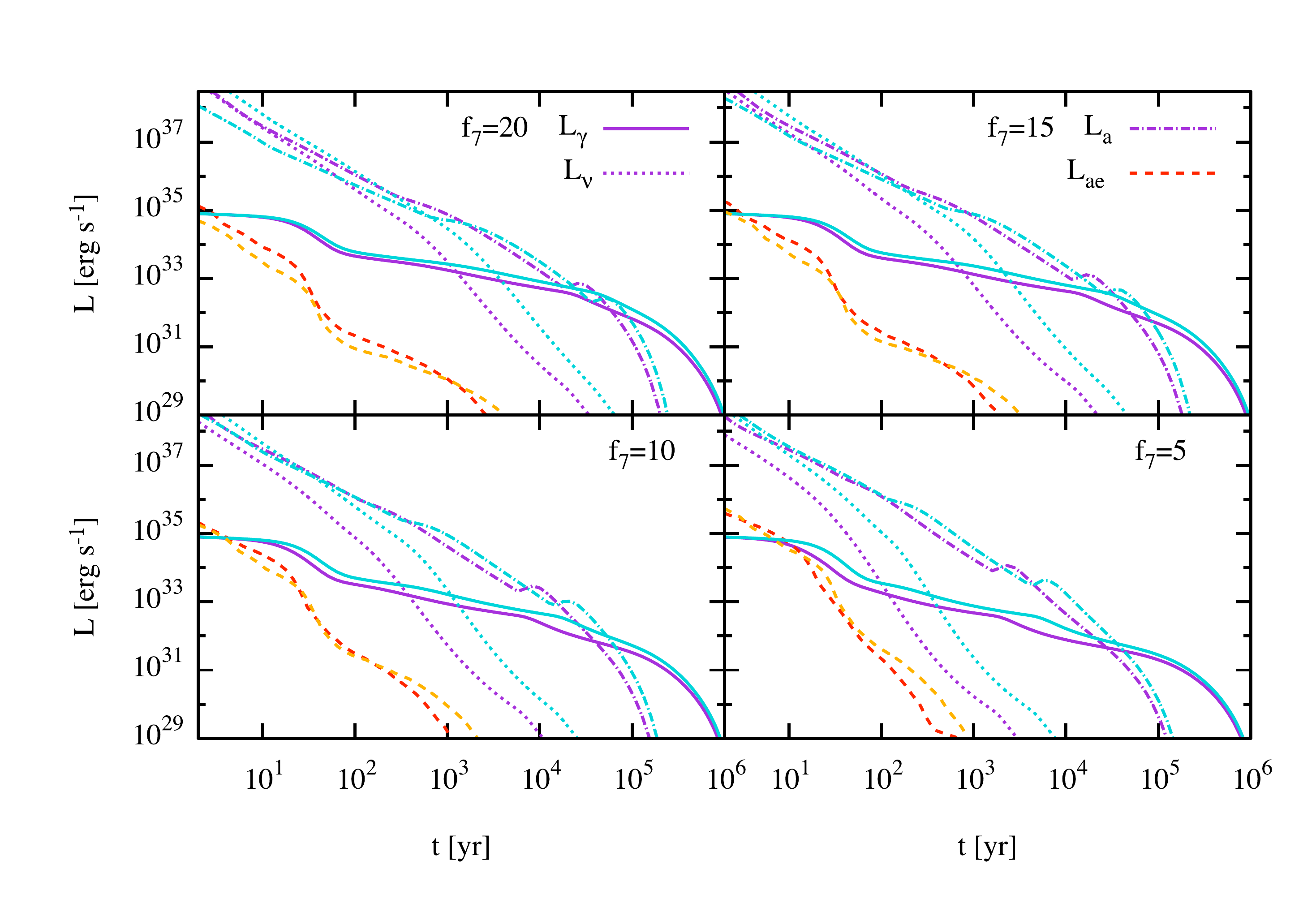}
    \caption{Dependence of the photon $L_\gamma$, neutrino $L_\nu$ and
      axion $L_a$ luminosities on age for the models of
      $M=1.4M_{\odot}$ stars for indicated values of the axion
      coupling $f_{a7}$. The PQ charges correspond to
      $\cos^2\beta=0.5$ (light blue) and $\cos^2\beta=1$ (violet). In
      addition we show the axion luminosity $L_{ae}$  due to electron
      bremsstrahlung in the crust [with emissivity given by
      Eq.~\eqref{eq:emissivity_brems}] for $\cos^2\beta=0.5$ (red) and
      $\cos^2\beta=1$ (orange).}
\label{fig:4}
\end{center}
\end{figure*}
\begin{figure*}
\begin{center}
\includegraphics[width=0.80\textwidth]{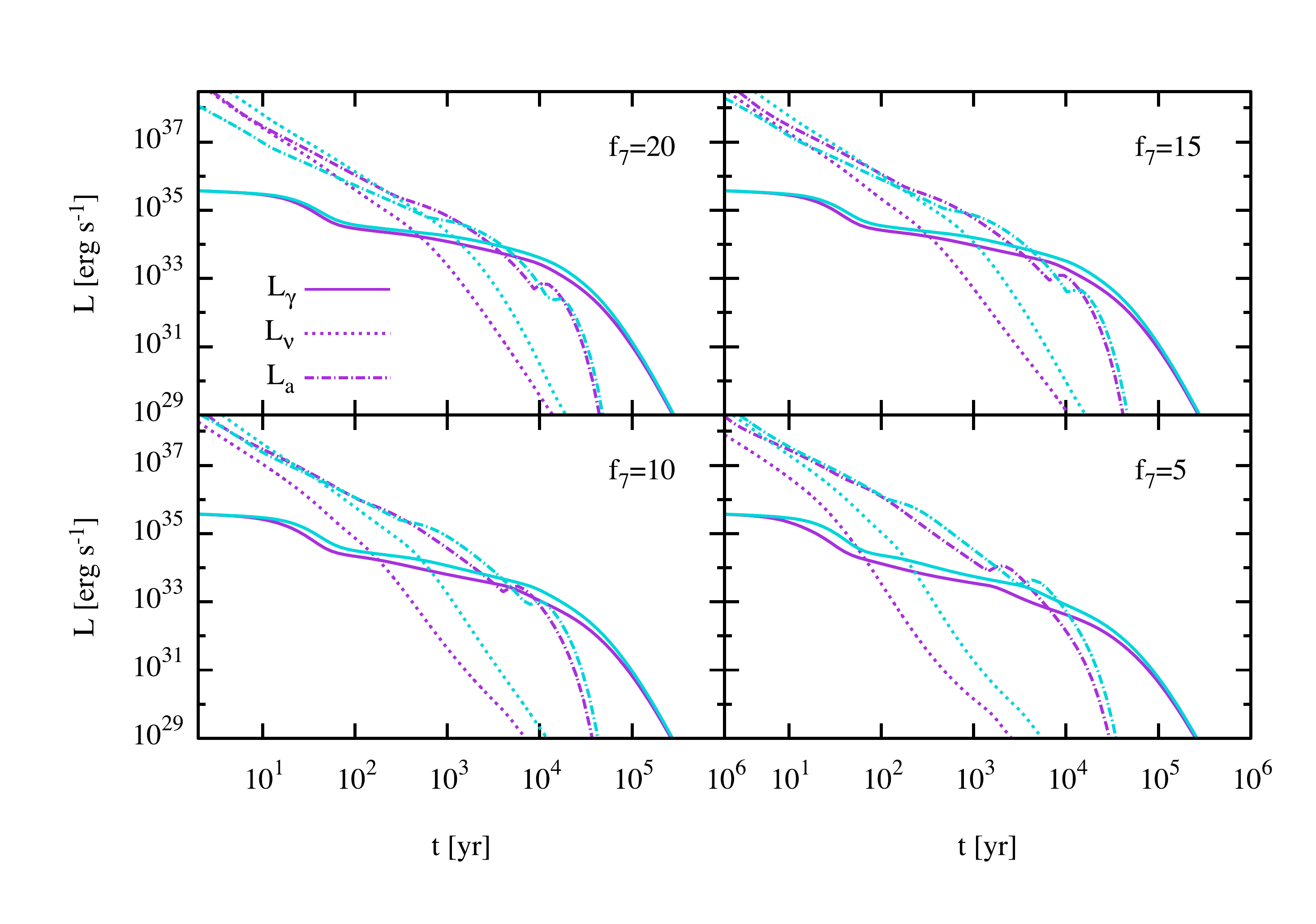}
\caption{
Same as in Fig.~\ref{fig:4} but for $\eta =1$, in which case the
photon luminosity is modified, but the neutrino and axion luminosities
are not. 
}
\label{fig:5}
\end{center}
\end{figure*}
\begin{figure}
\begin{center}
\includegraphics[width=0.49\textwidth]{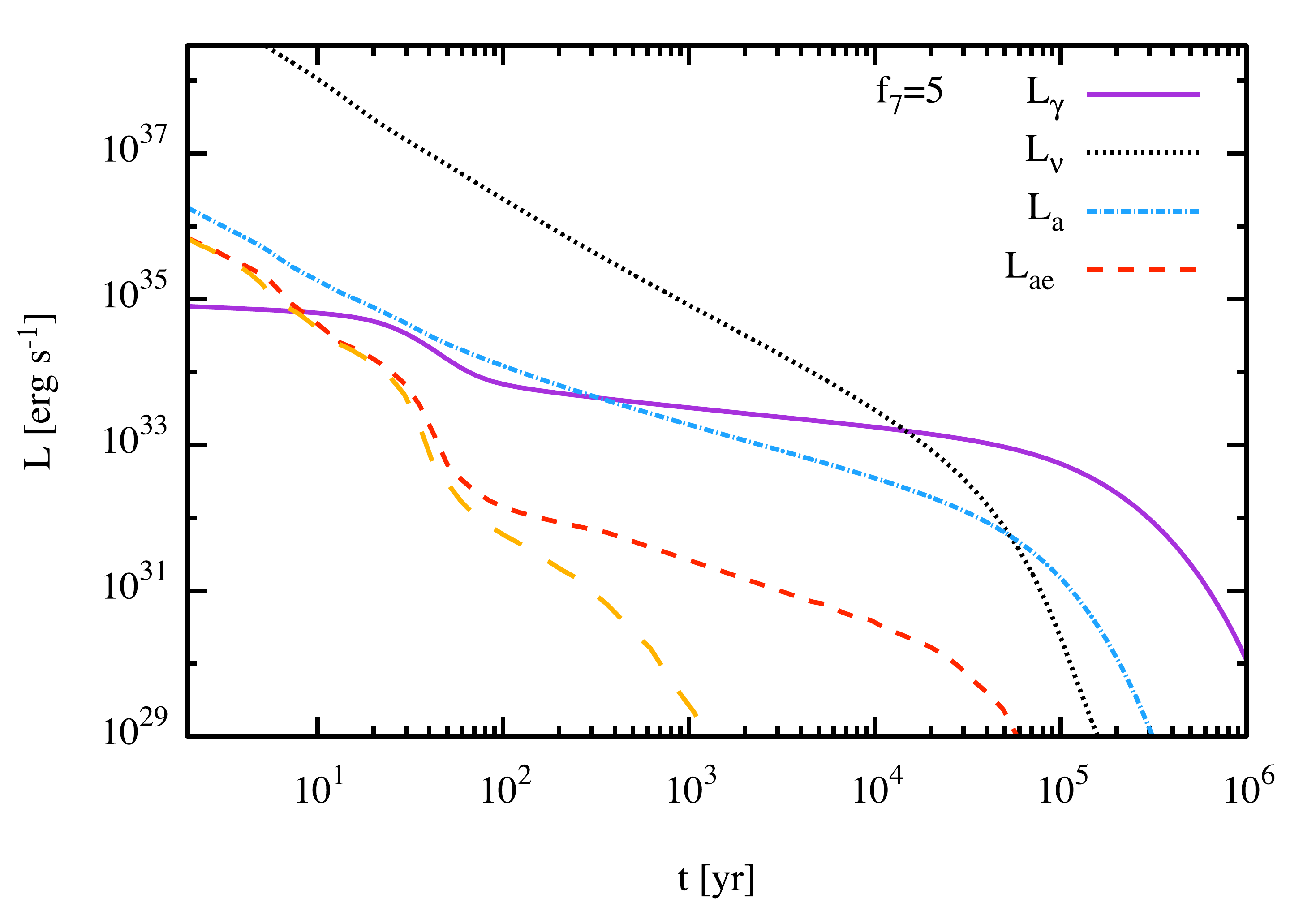}
\caption{ Axion and neutrino luminosities of a neutron star model with
  mass $M= 1.4M_{\odot}$ for the case of a nonaccreted iron envelope
  ($\eta = 0$) and $f_{a7}=2$.  We consider the value $\cos^2\beta = 0.344$ in which
  case $C_p = -0.284$, $C_e = 0.115$ and $C_n=0$ (neutrons do not
  couple to the axions).  The luminosity of axion bremsstrahlung by
  electrons $L_{ae}$ is shown for proton $^1S_0$ gap $\Delta_p = 0$
  (short dashed) and for gap values from Ref.~\cite{Takatsuka1973} 
  (long dashed). 
}
\label{fig:brem_v2}
\end{center}
\end{figure}
\begin{figure}[tbh]
\begin{center}
\includegraphics[width=0.45\textwidth]{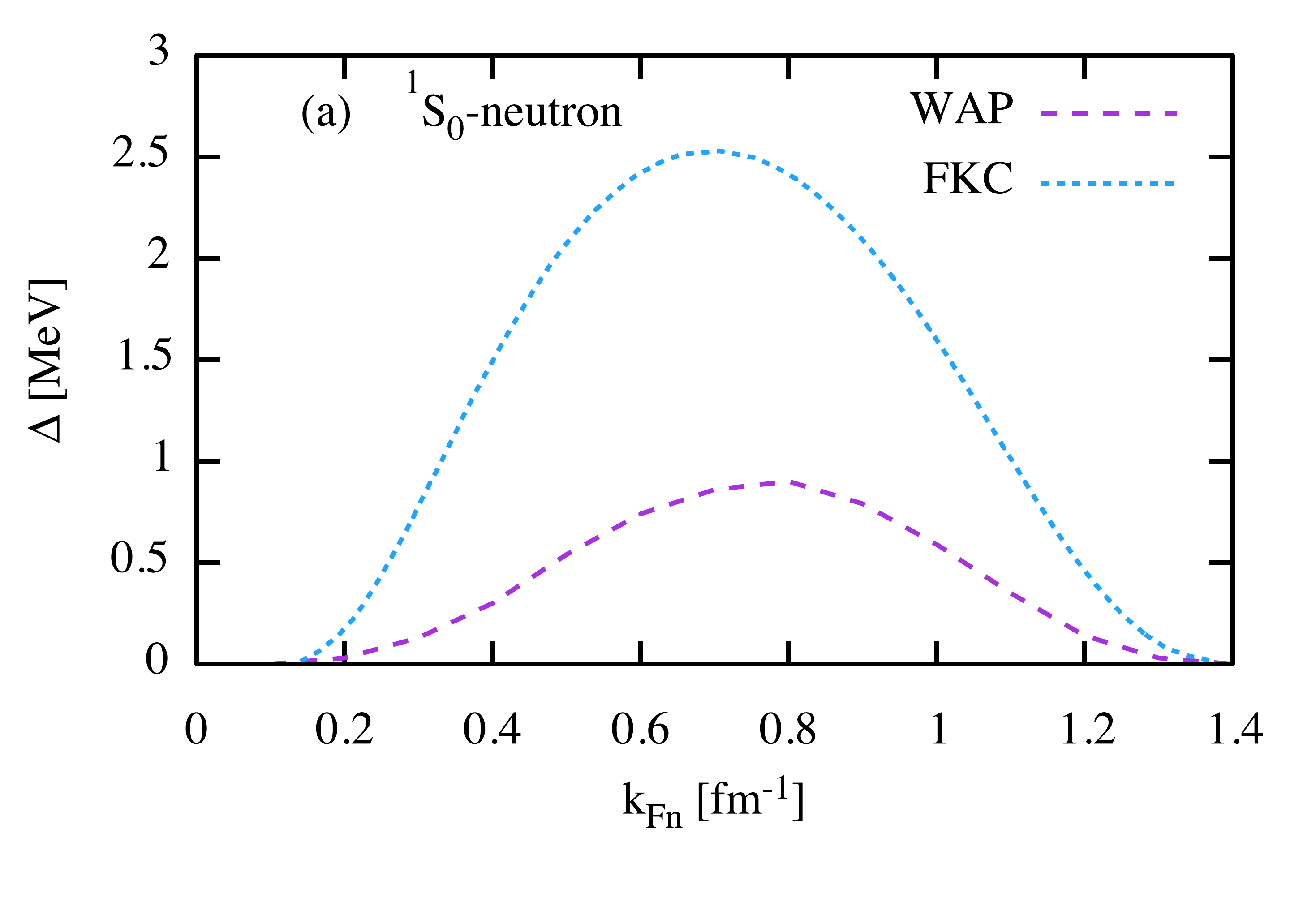}
\vskip -0.5cm
\includegraphics[width=0.45\textwidth]{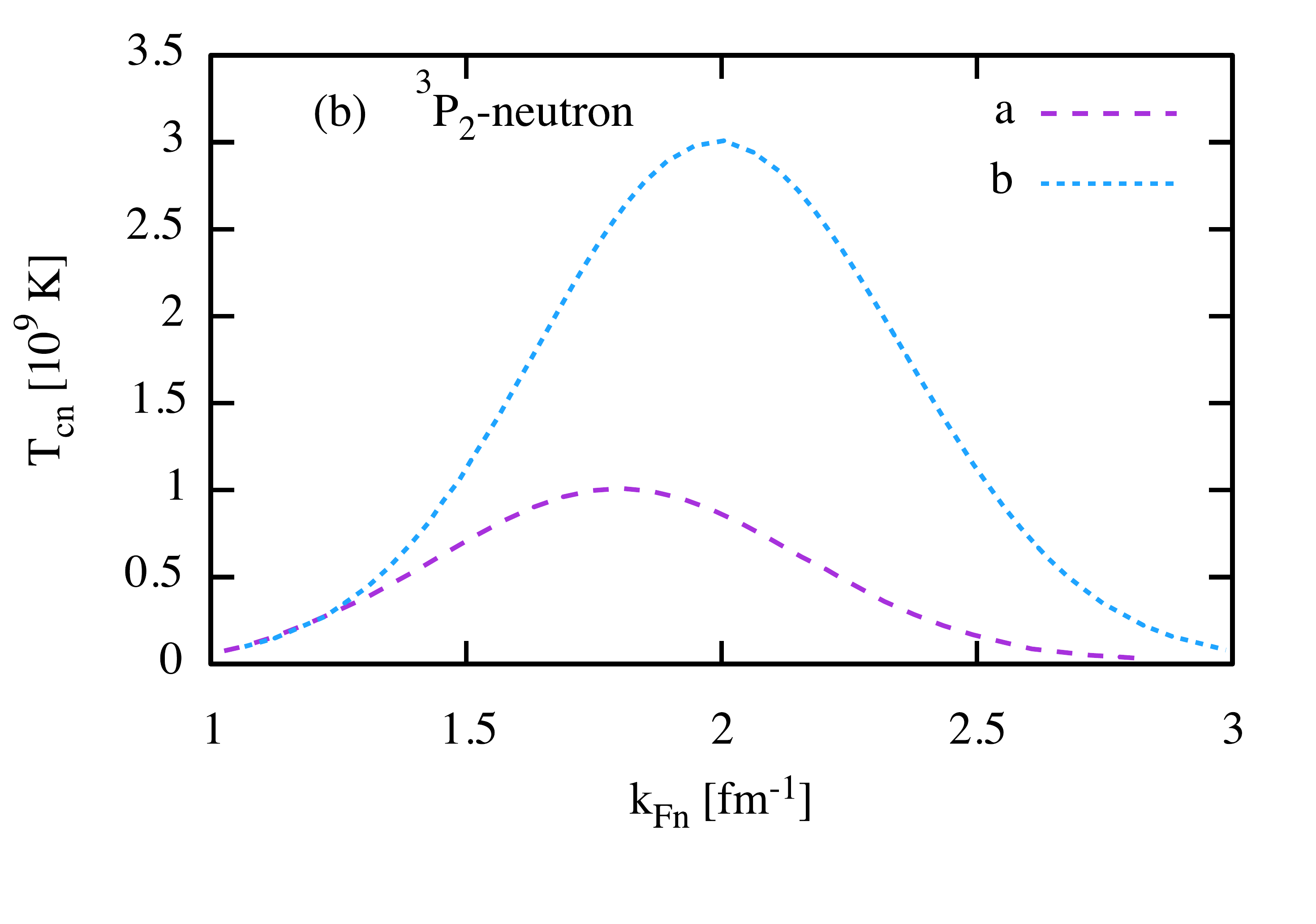}
\vskip -0.5cm
\includegraphics[width=0.45\textwidth]{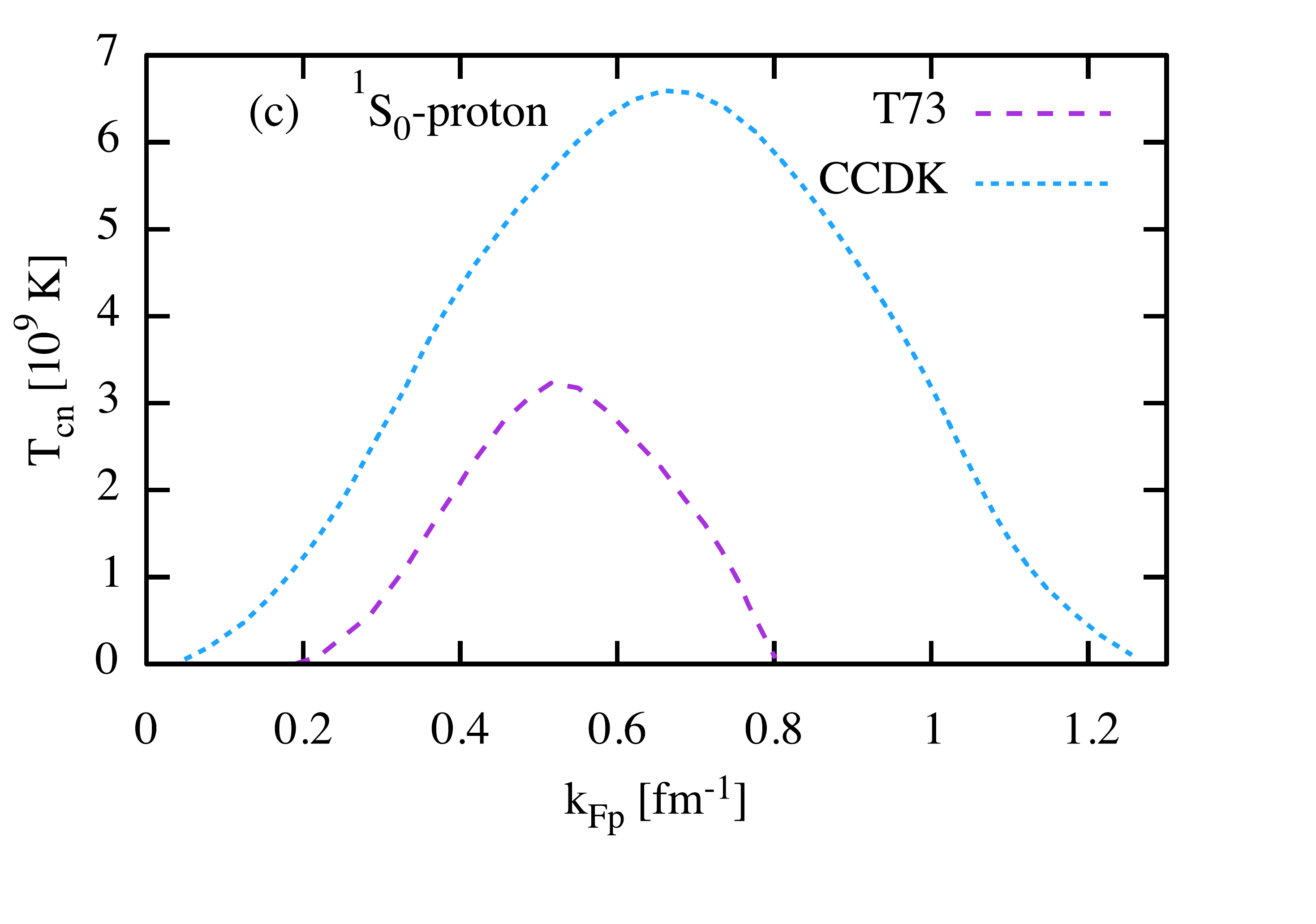}
\vskip -0.5cm
\caption{ 
Pairing gaps (critical temperatures) used in the simulations
  in Fig.~\ref{fig:gaps} as a function of neutron (proton) Fermi
  momentum.  (a) neutron $^1S_0$-gaps according to WAP~\cite{WAP1993}
  and FCK~\cite{FanKrotscheck2017}; (b) critical temperature of
  $^3P_2$-$^3F_2$ superfluid neutron phase transition according to
  models a and b of Ref.~\cite{2004ApJS..155..623P}; (c) critical
  temperature of $^1S_0$ superfluid proton phase transition according
  to T73~\cite{Takatsuka1973} and CCDK~\cite{CCDK1993}.  }
\label{fig:gaps2}
\end{center}
\end{figure}

\begin{figure*}[tbh]
\begin{center}
\includegraphics[width=0.80\textwidth]{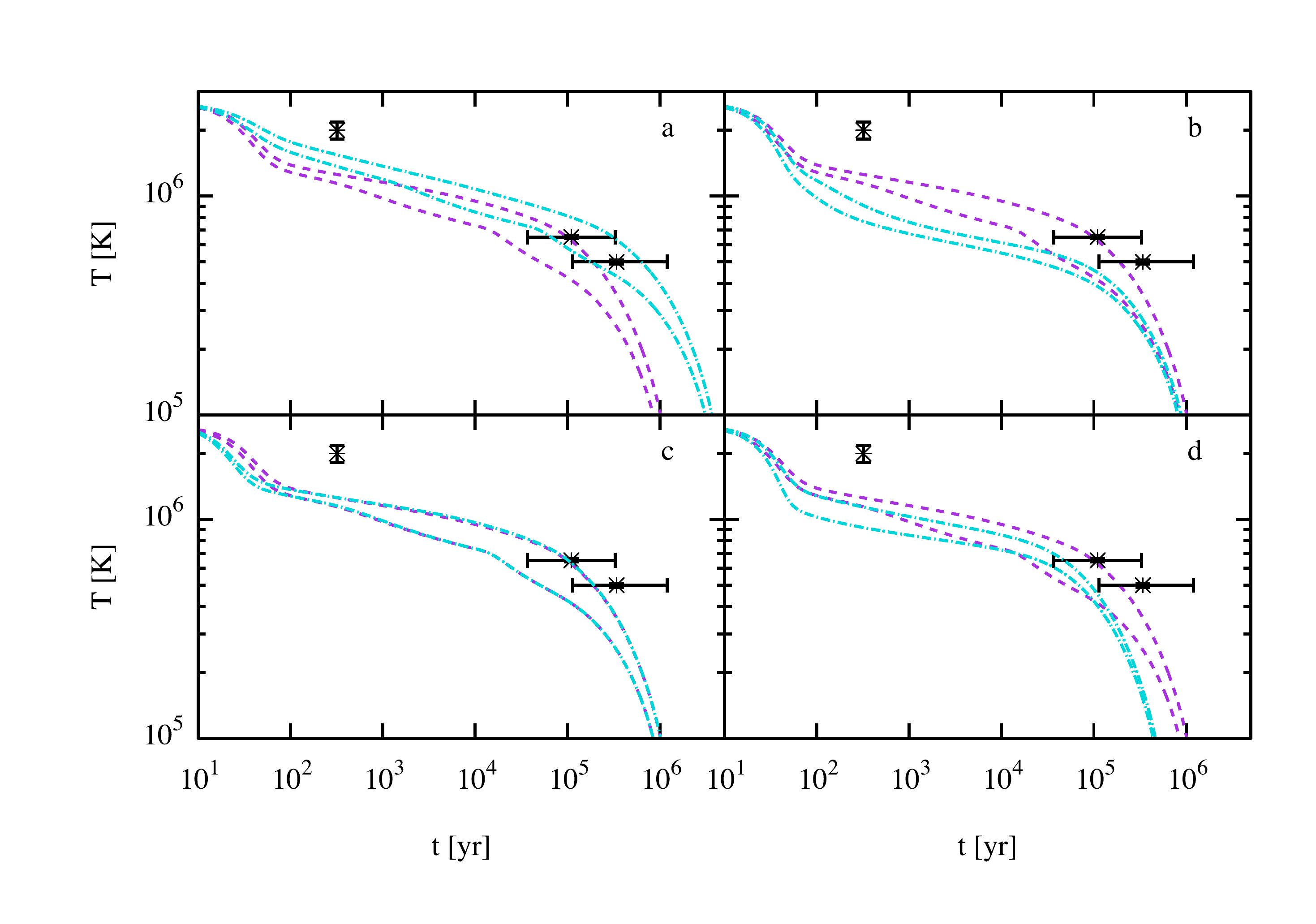}
\caption{ Cooling tracks of neutron star models with mass
  $M= 1.4M_{\odot}$ for the case of a nonaccreted iron envelope
  ($\eta = 0$), $f_{a7}=15$ and $\cos\beta =0$ and 1 (upper and lower
  curves, respectively) and for various selections of gaps, which are
  specified by triples of neutron $^1S_0$, neutron $^3P_2$-$^3F_2$,
  and proton $^1S_0$ gaps (see Fig.~\ref{fig:gaps2}).  Each panel
  features the results for the triple WAP-b-T73 (dashed lines) which
  is used in all models shown in Figs.~\ref{fig:2}-\ref{fig:brem_v2};
  the acronyms refer to Ref.~\cite{WAP1993} (WAP), model b of
  Ref.~\cite{2004ApJS..155..623P} (b), and Ref.~\cite{Takatsuka1973} (T73),
  respectively.  {\it Panel a:} Cooling tracks (dash-dotted lines) for
  gap triple WAP-0-T73, where 0 means vanishing neutron
  $^3P_2$-$^3F_2$ gap.  {\it Panel b:} Cooling tracks for an alternate
  model of $^3P_2$-$^3F_2$ gap specified by the triple WAP-a-T73
  (dash-dotted) lines; the acronym ``a'' correspond to pairing gap model
  a of Ref.~\cite{2004ApJS..155..623P}.  {\it Panel c:} Cooling tracks
  for an alternate model of neutron $^1S_0$ gap specified by the
  triple FKC-b-T73 dash-dotted lines, where the FKC refers to
  Ref.~\cite{FanKrotscheck2017}.  {\it Panel d:} Cooling tracks for proton
  $^1S_0$ gap specified by the triple WAP-b-CCDK (dash-dotted) line,
  where CCDK refers to Ref.~\cite{CCDK1993}.  }
\label{fig:gaps}
\end{center}
\end{figure*}

The dataset of surface temperatures considered in Paper I, which we
also use here is as follows. The first object -- the {\it CXO
  J232327.9+584842 } in the Cassiopeia A (Cas A) supernova remnant (SNR) --
is a representative of a group central compact objects (CCOs) --
pointlike, thermally emitting x-ray sources located close to the
geometrical centers of nonplerionic
SNRs~\cite{2013ApJ...765...58G}. These objects have low magnetic
fields, which exclude heating processes at this stage of evolution.
The value $T = 2.0\pm 0.18 \times 10^6$ K at the age $320$~yr was
used~\cite{2013ApJ...777...22E}. In addition three nearby neutron
stars which allow spectral fits to their x-ray emission were
considered~\cite{2005ApJ...623.1051D}. The fits invoke two black-body
temperatures and we identify the lowest one with the surface
temperature and quote only this value (see for further details Paper
I):
\begin{itemize}
\item {\it PSR B0656+14} with fit temperatures
  $T_w = (6.5\pm 0.1)\times 10^5$~K and characteristic age
  $1.1 \times 10^{2}$~kyr.
\item {\it PSR B1055-52} with  fit temperatures 
  $T_w = 7.9\pm 0.3\times 10^5$~K and characteristic age 
  $5.37 \times 10^{2}$~kyr. 
\item  {\it Geminga}, a radio-quiet object, with 
the  $T_w = 5.0\pm 0.1\times 10^5 $~K and 
characteristic age  $3.4 \times 10^{2}$ kyr.  
\end{itemize}
The error in the estimate of the ages of these objects from their
spin-down age is quantified by varying their age by a factor of 3. As
noted in Paper I, the data on PSR B1055-52 are marginally consistent
with the cooling curves even in the absence of axions. This can be
attributed to (a) larger error in the age of this pulsar than assumed
above; (b) internal heating; (c) the modeling of the pairing gaps,
which in principle can be tuned to fit the inferred temperature of PSR
B1055-52. Given the uncertainties involved, we will exclude the data on
PSR B1055-52 in the following. We do not attempt to fit the transient
behavior of the Cas A, as has been done in
Refs.~\cite{Leinson2014JCAP,Hamaguchi2018}, since the data on rapid
cooling is inconclusive~\cite{2013ApJ...777...22E,Posselt2018}. In any
case, the limits derived by these authors using cooling simulations
are comparable to those derived below. Future observations and
analysis of putative fast cooling of Cas A may prove to be a highly
efficient tool to constrain the properties of axions along the lines
of Refs.~\cite{Leinson2014JCAP,Hamaguchi2018}.  An additional
candidate for constraining axion properties is the peculiarly ``hot''
CCO HESS J1731-347. Reference~\cite{Beznogov2018} derived already limits on
$f_{a7}$ from cooling simulations using the data from this
object. Since this object challenges our understanding of the cooling
of neutron stars even without axionic cooling we will not include it
in our data set; see Ref.~\cite{Beznogov2018} for an alternative.

\subsection{Results of simulations}

A representative collection of 20 models of cooling neutron stars for
four values of the axion decay constant
$f_{a7} = 20, \, 15, \, 10, \, 5$ and the PQ changes specified by rows
1 to 5 in Table~\ref{tab:1} were simulated.  The mass of each model
was kept fixed at 1.4$M_{\odot}$ assuming the ``APR-Cat'' equation of
state of the {\sc NSCool} code. The triple of neutron $^1S_0$ and
$^3P_2$-$^3F_2$ and proton $^1S_0$ gaps were fixed to the value
``WAP-b-T73'' of the {\sc NSCool} code, where the acronyms refer to
Ref.~\cite{WAP1993} (WAP), model b of
Ref.~\cite{2004ApJS..155..623P} (b), and Ref.~\cite{Takatsuka1973}
(T73).  {We note that WAP and T73 gap values can be considered as
  lower bounds on the neutron and proton $^1S_0$ gaps
  respectively. Model b of Ref.~\cite{2004ApJS..155..623P} can be
  taken as an upper limit on the $^3P_2$-$^3F_2$ gap; we shall
  consider alternatives below. }

Figures~\ref{fig:2} and \ref{fig:3} show the results of cooling
simulations of 20 models of $m=M/M_{\odot}=1.4$ mass neutron stars
defined above with a nonaccreted iron envelope ($\eta = 0$) and a
light-element envelope ($\eta = 1$), respectively. Each of the panels
corresponds to a value of the axion coupling
$f_{a7} = 20, 15, 10$ and 5; within each panel, we vary the PQ charges
of neutrons, protons and electrons according to the indicated values
of $\cos^2\beta$ parameter. The dots with error bars show the three
test objects quoted above. Quite generally, the temperature of CCO in
Cas A is consistent with the cooling curves if one assumes a
light-element envelope in the absence of axion cooling; otherwise, its
theoretical temperature undershoots the observational value.  In
the case of older pulsars, the data agrees with the predictions of the
theoretical modeling without axion cooling only for an iron envelope.

Consider now switching on the axion production in the case $\eta = 0$
shown Fig.~\ref{fig:2}.  The additional loss of energy by axion
emission decreases the temperatures of our models.  For $f_{a7}=20$
all the five values of PQ charges are consistent with the data; for
$f_{a7}=15$ the values $\cos^2\beta=0.75$ and larger are excluded by
the data; for $f_{a7}=10$ the values larger than $\cos^2\beta = 0.5$ 
are incompatible with the data; finally, for $f_{a7}=5$ all values  of 
$\cos^2\beta$ are excluded by the data, except for    $\cos^2\beta=0.$

Similar, but not identical, conclusions are reached by examining the
data in Fig.~\ref{fig:3}. One observes that the following combinations
are inconsistent with the Cas A data: $f_{a7}=10 $ and
$\cos^2\beta=1.0$ and $f_{a7}=5$ and $\cos^2\beta \ge 0.25$. None of
the values of the PQ charges are excluded for $f_{a7}=15$ and 20.

Figures~\ref{fig:4} and ~\ref{fig:5} show the neutrino, axion and
photon luminosities as a function of time for four values of the axion
coupling constant $f_{a7}$ and PQ charges corresponding to
$\cos^2\beta=0.5$ and $\cos^2\beta=1$ (see Table~\ref{tab:1}) in the
cases $\eta=0$ and $\eta=1$, respectively. Clearly, the figures differ
only by the values of the surface photon luminosity, which is larger
in the case $\eta=1$ at early stages of thermal evolution and the
opposite is true at later stages of evolution.  It is seen that for
$f_{a7}=20$ the axion and neutrino luminosities are comparable. In the
remaining cases, the neutrino luminosity is subdominant and the
cooling rate is determined by the balance between the axion emission
rate and the change in the thermal energy given approximately by
$c_V dT/dt$, where $c_V$ is the net specific heat of the star. 

{To quantify the role of the electron bremsstrahlung of axions in
  the crust of a neutron star we show its luminosity $L_{ae}$ in
  Fig.~\ref{fig:4}. Irrespective of the value of $f_{a7}$ its
  contribution to axionic luminosity is negligibly small for
  $\cos^2\beta = 0.5$ and $1$; while its magnitude is comparable to
  the photon luminosity and even exceeds it for $f_{a7}\sim 5$, this
  occurs only at the early stages of evolution where the axion and
  neutrino emission by other processes dominate. However, the
  bremsstrahlung may contribute to the axion luminosity in some
  regions of the parameter space. Consider the case where the
  axion-neutron coupling $C_n =0$, which corresponds to
  $\cos^2\beta = 0.344$, see Fig.~\ref{fig:couplings}. In this case,
  the PBF processes on neutron condensates in the core and the crust
  do not contribute to axion emission. Then, the net axion luminosity
  is determined by the electron bremsstrahlung, the PBF process in the
  proton condensate, and proton modified Urca process. To disentangle
  the last two processes we consider two cases: (a) the proton gap
  vanishes in which case the proton PBF process vanishes as well; (b)
  the proton gap is finite and is given by
  T73~\cite{Takatsuka1973}. In case (a) the axionic cooling is the sum
  of the electron bremsstrahlung and modified Urca
  process. Figure~\ref{fig:brem_v2} shows that up to $\sim 10^2$ yr the
  electron bremsstrahlung can contribute a substantial fraction to the
  net axion luminosity in both cases (a) and (b). Of course, we
  consider only very special case of $C_n=0$; as seen in
  Fig.~\ref{fig:4} once neutrons couple to axions their axion
  emissivity completely  dominates the electron bremsstrahlung. Note
  that the magnitude of the axion luminosity as measured with respect
  to the neutrino luminosity changes with the value $f_{a7}$, whereas the
  relative magnitude of the luminosities of various axionic processes
  do not. This is a straightforward consequence of the $f_a^{-2}$
  scaling of the rates given by tree level amplitudes.}
 
{The modeling of neutron star cooling depends on a large number of
  parameters in general, but it is known to be most sensitive to the
  pairing gaps of neutrons and protons. To gain some insight in the
  effect of variation of these gaps, cooling simulations were
  performed with alternate pairing gaps for each type of the
  condensate while leaving the others fixed at their assumed values
  given by the triple WAP-b-T73 defined above and taken as a reference
  for comparison.  Figure~\ref{fig:gaps2} shows the pairing gaps or
  critical temperatures for neutron $^1S_0$ and $^3P_2$-$^3F_2$
  pairing and proton $^1S_0$ pairing. The reference gap in the neutron
  star crusts represents a lower limit (as it includes the suppression
  by long-range polarization effects). As an alternative we use the
  computation of Ref.~\cite{FanKrotscheck2017} where the pairing
  interaction resums both long- and short-range correlation in an
  approximated way. The resulting gap is significantly larger, the
  maxima differing by a factor $\sim 2.5$. The reference value of the
  critical temperature for neutron $^3P_2$-$^3F_2$ pairing is large and
  we need to adopt a smaller value; we consider below two options where
  $T_c = 0$ in this channel or it is given by the model a (instead
  of b) of Ref.~\cite{2004ApJS..155..623P}. Finally, for proton
  $^1S_0$ pairing we explore the possibility of larger $T_c$, taking
  as an example the one corresponding to the gap given by
  Ref.~\cite{CCDK1993}. The corresponding critical temperature has a
  maximum by a factor of 2 larger and a substantially larger extent into
  the core of the star.

  We start with the case of vanishing neutron $^3P_2$-$^3F_2$ gap
  [panel (a) of Fig.~\ref{fig:gaps}].  This results in moderately
  enhanced  temperatures, the uncertainty being of the order
  of $5\%$ except during the late time cooling $t> 10^5$ yr where
  significant differences arise.  Adopting a smaller value of the
  $^3P_2$-$^3F_2$ gap [panel (b)] we find that cooling tracks drop
  earlier to lower temperatures at $t\sim 10^2$ yr and the
  temperatures stay lower throughout the neutrino-axion cooling era
  $t\le 10^5$ yr. This implies that lower values of the pairing
  $^3P_2$-$^3F_2$ gap cannot affect the limits inferred using its
  reference value.  Employing a larger neutron $^1S_0$ pairing
  gap~\cite{FanKrotscheck2017} in the crusts [panel (c)] leads to an
  earlier drop in the cooling curves at $10^2$ yr and temperatures
  beyond this timescale almost identical to the reference ones; this
  in turn implies that the inferred limits will not be affected with
  the variations of the neutron $^1S_0$ gap. Finally, if one adopts a
  larger proton $^1S_0$ gap~\cite{CCDK1993} (panel d) the drop in the
  cooling curves is larger at $t\le 10^2$ yr. In this case the
  deviations are again not large, of the order of $10\%$ for
  $t\le 10^5$ yr .  We conclude that the variations in the values of
  the gaps in neutron and proton condensates do not affect
  significantly the limits drawn from the analysis of the cooling
  curves. }

In the case of Cas A the age of the CCO is known, therefore its
average temperature provides a reliable reference value. In the
parameter space spanned by 
$\cos\beta^2$ and $f_{a7}$ we can now deduces
the limiting values of these parameters. As seen from Fig.~\ref{fig:3}
in the range $0\le \cos\beta^2\le 1$ the compatibility of the data implies
$5\le f_{a7}\le 10$. Using Eq.~\eqref{eq:axion_mass} we find upper
limits on axion masses
\bea
&&\textrm {max } [m_a] \simeq 0.12~\textrm{eV}, \quad \cos^2\beta
\simeq 0,\\
&&\textrm {max } [m_a] \simeq 0.06~\textrm{eV}, \quad \cos^2\beta \simeq 1.
\eea
Thus, the DFSZ axion mass above the quoted values is excluded by
numerical simulations of cooling neutron stars and their comparison
with the observational data on Cas A.

\begin{figure}[tbh]
\begin{center}
\includegraphics[width=0.45\textwidth]{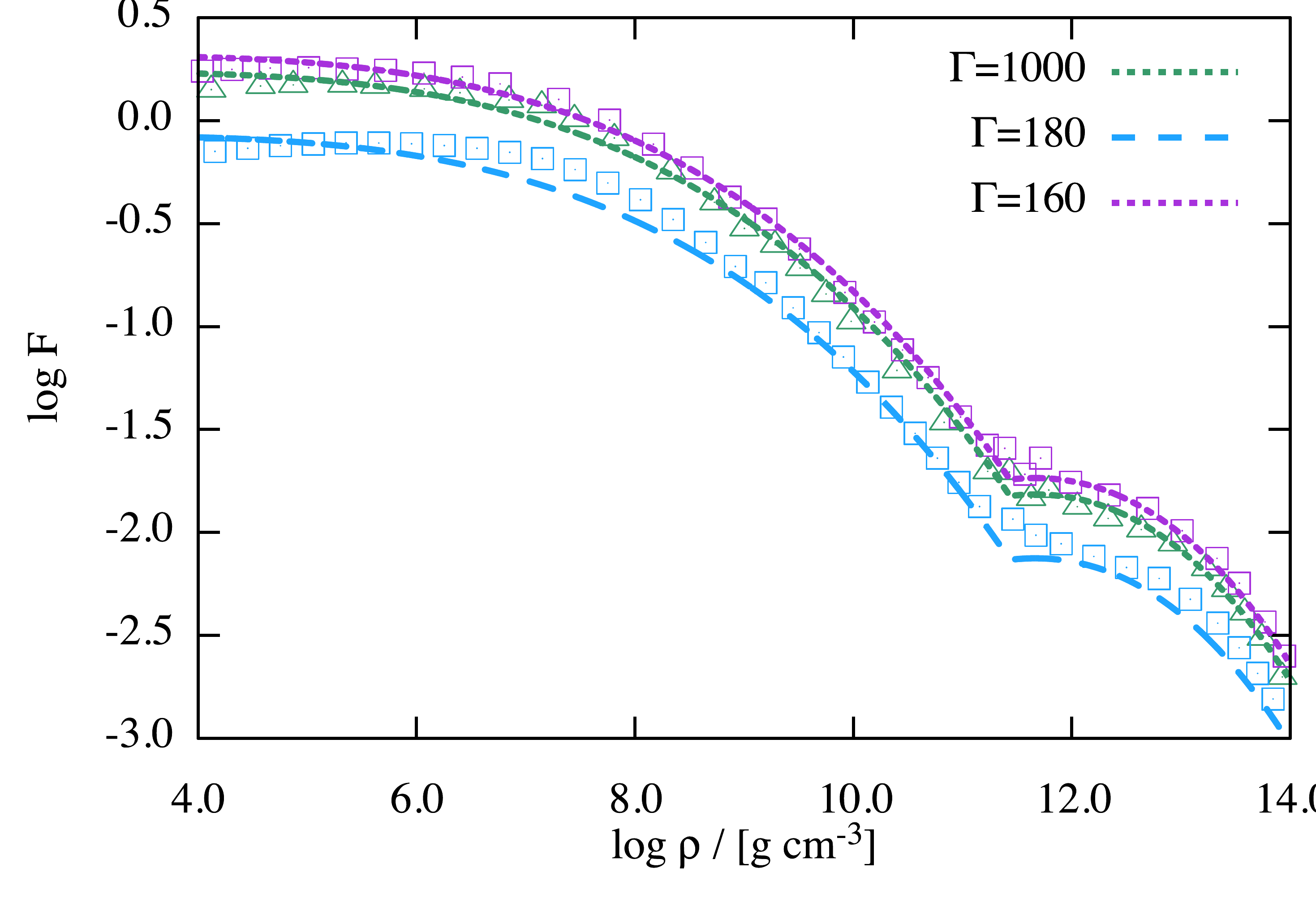}
\caption{ Computations of Ref.~\cite{Nakagawa1987} (points) are
  compared with the fits (lines) for the indicated values of the
  parameter $\Gamma$.  }
\label{fig:Fcorr}
\end{center}
\end{figure}

\section{Discussion and conclusions}
\label{sec:conclusions}

In this work, we continued our study of cooling of weakly magnetized
neutron stars by the emission of axions. The key strategy (see also
Paper I) is to assume that the observed objects are not heavy enough
to allow for nucleation of new degrees of freedom in their
high-density cores. This cooling behavior could be changed
significantly in cases of quark matter
nucleation~\cite{2001A&A...368..561B,2011PhRvD..84f3015H,
  2005PhRvD..71k4011A,2016EPJA...52...44S} or
hyperonization~\cite{2018MNRAS.475.4347R,2018arXiv180801819G,2018ApJ...863..104N};
for reviews see
Refs.~\cite{weber_book,2007PrPNP..58..168S,Page2013,2018arXiv180200017S}.
Even without new degrees of freedom, high densities may permit fast
(or accelerated) processes involving only neutrons, protons and
leptons~\cite{2004A&A...424..979B,2012PhRvC..85b2802B}. For canonical
mass stars with masses $M\sim 1.4M_{\odot}$ neutrino cooling is
slow~\cite{2009ApJ...707.1131P,2018PhRvC..97a5804B}. Taking the
consistency of the neutrino cooling models with the data on three
neutron stars with reliable fits to their blackbody emission as a
reference point, we explored the modification introduced by switching
on the axion emission from the stellar interior. We have included all
the relevant axion emission processes which couple axions to
electrons, protons and neutrons, in particular, the recently derived
rates from PBF processes~\cite{2013NuPhA.897...62K,Sedrakian2016}, as
well as axion emission by electron
bremsstrahlung~\cite{Nakagawa1987,Nakagawa1988}.  {The last
  process of electron bremsstrahlung is an insignificant source of
  axion emission, except when the neutron PQ charge is extremely
  small, so that the coupling of neutrons to axions can be neglected.}

In this work, we focused on the DFSZ model which allows axion emission
from the electronic component of the star. The DFSZ model has the
advantage that the PQ charges of hadrons and electrons are locked via
a single parameter $\cos^2\beta$. The limiting value of the axion
coupling constant then spans a wide range $5\le f_{a7}\le 15$
depending on the value of $\cos^2\beta$. This translates into a range
of upper bounds on the axion mass
\bea 
\label{eq:axion_mass2}
 0.06\le {\rm max}\, [m_a] \le 0.12~ \textrm{eV},
\eea
which are consistent with those inferred for the  KSVZ model.

\section*{ACKNOWLEDGMENTS} The support by the Deutsche
Forschungsgemeinschaft (Grants No. SE 1836/3-2 and  SE 1836/4-1)
is gratefully acknowledged. Partial support was provided by the
European COST Actions ``NewCompStar'' (MP1304), ``PHAROS'' (CA16214),
and the State of Hesse LOEWE-Program in HIC for FAIR.

\appendix

\section{Fits to the correlation functions $F_{L/S}$}

The correlation functions $F_{L/S}$ in Eq.~\eqref{eq:emissivity_brems}
have been computed in Ref.~\cite{Nakagawa1987}. We used the following fit
formulas for these functions to implement the axion bremsstrahlung by
electrons. As a function of the density these are given by simple
polynomials 
\bea
\log F_{S/L}(x,\Gamma) =  a + b x^2 +  cx^4 - [1 -u_{S/L} (\Gamma)],
\eea
with $x \equiv\log \rho$ where 
\bea 
 a = 0.21946, \quad b = 0.00287263, \quad c = -0.000142016,\nonumber\\
\eea
for $x\le x_0  = 11.4$, and 
\bea 
a = -6.47808, \quad  b = 0.068645, \quad  c = -0.000252677,\nonumber\\
\eea
for $x > x_0 $.  These fits were carried out for $\Gamma_0 = 10^3$ and
subsequently extrapolated to other relevant values of $\Gamma$ using
the function
\bea
u_{S/L} = u_0 +  u_1 (\Gamma/\Gamma_0) +u_2 (\Gamma/\Gamma_0)^2,
\eea
where
\bea 
u_0 = 0.488049, \quad u_1= 1.25585, \quad  u_2 = -0.743902, \nonumber\\
\eea
in the solid phase and 
\bea 
u_0 = 0.672409, \quad u_1= 0.182774, \quad  u_2 = 0.144817, \nonumber\\
\eea
in the liquid phase.  Figure~\ref{fig:Fcorr} shows the dependence of the
correlation functions on the density on a log-log plot. The overall accuracy of
the fit is below $10\%$, with some larger deviations $\le 30\%$ in
a narrow range of densities for selected values of $\Gamma.$


\begin{thebibliography}{65}%
\makeatletter
\providecommand \@ifxundefined [1]{%
 \@ifx{#1\undefined}
}%
\providecommand \@ifnum [1]{%
 \ifnum #1\expandafter \@firstoftwo
 \else \expandafter \@secondoftwo
 \fi
}%
\providecommand \@ifx [1]{%
 \ifx #1\expandafter \@firstoftwo
 \else \expandafter \@secondoftwo
 \fi
}%
\providecommand \natexlab [1]{#1}%
\providecommand \enquote  [1]{``#1''}%
\providecommand \bibnamefont  [1]{#1}%
\providecommand \bibfnamefont [1]{#1}%
\providecommand \citenamefont [1]{#1}%
\providecommand \href@noop [0]{\@secondoftwo}%
\providecommand \href [0]{\begingroup \@sanitize@url \@href}%
\providecommand \@href[1]{\@@startlink{#1}\@@href}%
\providecommand \@@href[1]{\endgroup#1\@@endlink}%
\providecommand \@sanitize@url [0]{\catcode `\\12\catcode `\$12\catcode
  `\&12\catcode `\#12\catcode `\^12\catcode `\_12\catcode `\%12\relax}%
\providecommand \@@startlink[1]{}%
\providecommand \@@endlink[0]{}%
\providecommand \url  [0]{\begingroup\@sanitize@url \@url }%
\providecommand \@url [1]{\endgroup\@href {#1}{\urlprefix }}%
\providecommand \urlprefix  [0]{URL }%
\providecommand \Eprint [0]{\href }%
\providecommand \doibase [0]{http://dx.doi.org/}%
\providecommand \selectlanguage [0]{\@gobble}%
\providecommand \bibinfo  [0]{\@secondoftwo}%
\providecommand \bibfield  [0]{\@secondoftwo}%
\providecommand \translation [1]{[#1]}%
\providecommand \BibitemOpen [0]{}%
\providecommand \bibitemStop [0]{}%
\providecommand \bibitemNoStop [0]{.\EOS\space}%
\providecommand \EOS [0]{\spacefactor3000\relax}%
\providecommand \BibitemShut  [1]{\csname bibitem#1\endcsname}%
\let\auto@bib@innerbib\@empty
\bibitem [{\citenamefont {{Wilczek}}(1978)}]{1978PhRvL..40..279W}%
  \BibitemOpen
  \bibfield  {author} {\bibinfo {author} {\bibfnamefont {F.}~\bibnamefont
  {{Wilczek}}},\ }\href {\doibase 10.1103/PhysRevLett.40.279} {\bibfield
  {journal} {\bibinfo  {journal} {\prl}\ }\textbf {\bibinfo {volume} {40}},\
  \bibinfo {pages} {279} (\bibinfo {year} {1978})}\BibitemShut {NoStop}%
\bibitem [{\citenamefont {{Weinberg}}(1978)}]{1978PhRvL..40..223W}%
  \BibitemOpen
  \bibfield  {author} {\bibinfo {author} {\bibfnamefont {S.}~\bibnamefont
  {{Weinberg}}},\ }\href {\doibase 10.1103/PhysRevLett.40.223} {\bibfield
  {journal} {\bibinfo  {journal} {\prl}\ }\textbf {\bibinfo {volume} {40}},\
  \bibinfo {pages} {223} (\bibinfo {year} {1978})}\BibitemShut {NoStop}%
\bibitem [{\citenamefont {{'t Hooft}}(1976)}]{1976PhRvL..37....8T}%
  \BibitemOpen
  \bibfield  {author} {\bibinfo {author} {\bibfnamefont {G.}~\bibnamefont {{'t
  Hooft}}},\ }\href {\doibase 10.1103/PhysRevLett.37.8} {\bibfield  {journal}
  {\bibinfo  {journal} {\prl}\ }\textbf {\bibinfo {volume} {37}},\ \bibinfo
  {pages} {8} (\bibinfo {year} {1976})}\BibitemShut {NoStop}%
\bibitem [{\citenamefont {{Peccei}}\ and\ \citenamefont
  {{Quinn}}(1977)}]{1977PhRvL..38.1440P}%
  \BibitemOpen
  \bibfield  {author} {\bibinfo {author} {\bibfnamefont {R.~D.}\ \bibnamefont
  {{Peccei}}}\ and\ \bibinfo {author} {\bibfnamefont {H.~R.}\ \bibnamefont
  {{Quinn}}},\ }\href {\doibase 10.1103/PhysRevLett.38.1440} {\bibfield
  {journal} {\bibinfo  {journal} {\prl}\ }\textbf {\bibinfo {volume} {38}},\
  \bibinfo {pages} {1440} (\bibinfo {year} {1977})}\BibitemShut {NoStop}%
\bibitem [{\citenamefont {{Peccei}}(2008)}]{2008LNP...741....3P}%
  \BibitemOpen
  \bibfield  {author} {\bibinfo {author} {\bibfnamefont {R.~D.}\ \bibnamefont
  {{Peccei}}},\ }in\ \href@noop {} {\emph {\bibinfo {booktitle} {Axions}}},\
  \bibinfo {editor} {edited by\ \bibinfo {editor}
  {\bibfnamefont {M.}~\bibnamefont {{Kuster}}}, \bibinfo {editor}
  {\bibfnamefont {G.}~\bibnamefont {{Raffelt}}}, \ and\ \bibinfo {editor}
  {\bibfnamefont {B.}~\bibnamefont {{Beltr{\'a}n}}}}\ (\bibinfo {year}
{Springer Verlag, Berlin, 2008), Vol. 741, }\
  pp.\ \bibinfo {pages} {3--540.}\ \Eprint
  {http://arxiv.org/abs/hep-ph/0607268} {hep-ph/0607268} \BibitemShut {NoStop}%
\bibitem [{\citenamefont {{Ringwald}}(2012)}]{Ringwald2012}%
  \BibitemOpen
  \bibfield  {author} {\bibinfo {author} {\bibfnamefont {A.}~\bibnamefont
  {{Ringwald}}},\ }\href {\doibase 10.1016/j.dark.2012.10.008} {\bibfield
  {journal} {\bibinfo  {journal} {Phys. Dark Universe}\ }\textbf
  {\bibinfo {volume} {1}},\ \bibinfo {pages} {116} (\bibinfo {year} {2012})},\
  \Eprint {http://arxiv.org/abs/1210.5081} {arXiv:1210.5081 [hep-ph]}
  \BibitemShut {NoStop}%
\bibitem [{\citenamefont {{Giannotti}}\ \emph
  {et~al.}(2016{\natexlab{a}})\citenamefont {{Giannotti}}, \citenamefont
  {{Irastorza}}, \citenamefont {{Redondo}}, \ and\ 
  \citenamefont {{Ringwald}}}]{Giannotti2016}%
  \BibitemOpen
  \bibfield  {author} {\bibinfo {author} {\bibfnamefont {M.}~\bibnamefont
  {{Giannotti}}}, \bibinfo {author} {\bibfnamefont {I.~G.}\ \bibnamefont
  {{Irastorza}}}, \bibinfo {author} {\bibfnamefont {J.}~\bibnamefont
  {{Redondo}}}, \ and\ \bibinfo {author} {\bibfnamefont {A.}~\bibnamefont
  {{Ringwald}}}, \ }\href {\doibase 10.1088/1475-7516/2017/10/010} {\bibfield
  {journal} {\bibinfo  {journal} {\jcap}\ }\textbf {\bibinfo {volume} {05}},\
  \bibinfo {eid} {057} (\bibinfo {year} {2016}{\natexlab{a}})}\BibitemShut
  {NoStop}
\bibitem [{\citenamefont {{Irastorza}}\ and\ \citenamefont
  {{Redondo}}(2018)}]{Irastorza2018}%
  \BibitemOpen
  \bibfield  {author} {\bibinfo {author} {\bibfnamefont {I.~G.}\ \bibnamefont
  {{Irastorza}}}\ and\ \bibinfo {author} {\bibfnamefont {J.}~\bibnamefont
  {{Redondo}}},\ }\href {\doibase 10.1016/j.ppnp.2018.05.003} {\bibfield
  {journal} {\bibinfo  {journal} {Prog. Part. Nucl. Phys.}\ }\textbf {\bibinfo
  {volume} {102}},\ \bibinfo {pages} {89} (\bibinfo {year} {2018})},\ \Eprint
  {http://arxiv.org/abs/1801.08127} {arXiv:1801.08127 [hep-ph]} \BibitemShut
  {NoStop}%
\bibitem [{\citenamefont {{Sedrakian}}(2016{\natexlab{a}})}]{Sedrakian2016}%
  \BibitemOpen
  \bibfield  {author} {\bibinfo {author} {\bibfnamefont {A.}~\bibnamefont
  {{Sedrakian}}},\ }\href {\doibase 10.1103/PhysRevD.93.065044} {\bibfield
  {journal} {\bibinfo  {journal} {\prd}\ }\textbf {\bibinfo {volume} {93}},\
  \bibinfo {eid} {065044} (\bibinfo {year} {2016}{\natexlab{a}})},\ \Eprint
  {http://arxiv.org/abs/1512.07828} {arXiv:1512.07828 [astro-ph.HE]}
  \BibitemShut {NoStop}%
\bibitem [{\citenamefont {{Raffelt}}\ \emph {et~al.}(2011)\citenamefont
  {{Raffelt}}, \citenamefont {{Redondo}},\ and\ \citenamefont
  {{Maira}}}]{2011PhRvD..84j3008R}%
  \BibitemOpen
  \bibfield  {author} {\bibinfo {author} {\bibfnamefont {G.~G.}\ \bibnamefont
  {{Raffelt}}}, \bibinfo {author} {\bibfnamefont {J.}~\bibnamefont
  {{Redondo}}}, \ and\ \bibinfo {author} {\bibfnamefont {N.~V.}\ \bibnamefont
  {{Maira}}},\ }\href {\doibase 10.1103/PhysRevD.84.103008} {\bibfield
  {journal} {\bibinfo  {journal} {\prd}\ }\textbf {\bibinfo {volume} {84}},\
  \bibinfo {eid} {103008} (\bibinfo {year} {2011})},\ \Eprint
  {http://arxiv.org/abs/1110.6397} {arXiv:1110.6397 [hep-ph]} \BibitemShut
  {NoStop}%
\bibitem [{\citenamefont {{Giannotti}}\ \emph
  {et~al.}(2017{\natexlab{a}})\citenamefont {{Giannotti}}, \citenamefont
  {{Irastorza}}, \citenamefont {{Redondo}}, \citenamefont {{Ringwald}},\ and\
  \citenamefont {{Saikawa}}}]{Giannotti2017}%
  \BibitemOpen
  \bibfield  {author} {\bibinfo {author} {\bibfnamefont {M.}~\bibnamefont
  {{Giannotti}}}, \bibinfo {author} {\bibfnamefont {I.~G.}\ \bibnamefont
  {{Irastorza}}}, \bibinfo {author} {\bibfnamefont {J.}~\bibnamefont
  {{Redondo}}}, \bibinfo {author} {\bibfnamefont {A.}~\bibnamefont
  {{Ringwald}}}, \ and\ \bibinfo {author} {\bibfnamefont {K.}~\bibnamefont
  {{Saikawa}}},\ }\href {\doibase 10.1088/1475-7516/2017/10/010} {\bibfield
  {journal} {\bibinfo  {journal} {\jcap}\ }\textbf {\bibinfo {volume} {10}},\
  \bibinfo {eid} {010} (\bibinfo {year} {2017}{\natexlab{a}})},\ \Eprint
  {http://arxiv.org/abs/1708.02111} {arXiv:1708.02111 [hep-ph]} \BibitemShut
  {NoStop}%
\bibitem [{\citenamefont {{Kim}}(1979)}]{Kim1979}%
  \BibitemOpen
  \bibfield  {author} {\bibinfo {author} {\bibfnamefont {J.~E.}\ \bibnamefont
  {{Kim}}},\ }\href {\doibase 10.1103/PhysRevLett.43.103} {\bibfield  {journal}
  {\bibinfo  {journal} {\prl}\ }\textbf {\bibinfo {volume} {43}},\ \bibinfo
  {pages} {103} (\bibinfo {year} {1979})}\BibitemShut {NoStop}%
\bibitem [{\citenamefont {Shifman}\ \emph {et~al.}(1980)\citenamefont
  {Shifman}, \citenamefont {Vainshtein},\ and\ \citenamefont
  {Zakharov}}]{Shiftman1980}%
  \BibitemOpen
  \bibfield  {author} {\bibinfo {author} {\bibfnamefont {M.}~\bibnamefont
  {Shifman}}, \bibinfo {author} {\bibfnamefont {A.}~\bibnamefont {Vainshtein}},
  \ and\ \bibinfo {author} {\bibfnamefont {V.}~\bibnamefont {Zakharov}},\
  }\href@noop {} {\bibfield  {journal} {\bibinfo  {journal} {\nphysb}\ }\textbf
  {\bibinfo {volume} {B166}},\ \bibinfo {pages} {493 } (\bibinfo {year}
  {1980})}\BibitemShut {NoStop}%
\bibitem [{\citenamefont {Dine}\ \emph {et~al.}(1981)\citenamefont {Dine},
  \citenamefont {Fischler},\ and\ \citenamefont {Srednicki}}]{Dine1981}%
  \BibitemOpen
  \bibfield  {author} {\bibinfo {author} {\bibfnamefont {M.}~\bibnamefont
  {Dine}}, \bibinfo {author} {\bibfnamefont {W.}~\bibnamefont {Fischler}}, \
  and\ \bibinfo {author} {\bibfnamefont {M.}~\bibnamefont {Srednicki}},\
  }\href@noop {} {\bibfield  {journal} {\bibinfo  {journal} {Phys. Lett. B}\
  }\textbf {\bibinfo {volume} {104B}},\ \bibinfo {pages} {199 } (\bibinfo {year}
  {1981})}\BibitemShut {NoStop}%
\bibitem [{\citenamefont {Zhitnitsky}(1980)}]{Zhitnitsky:1980tq}%
  \BibitemOpen
  \bibfield  {author} {\bibinfo {author} {\bibfnamefont {A.~R.}\ \bibnamefont
  {Zhitnitsky}},\ }\href@noop {} {\bibfield  {journal} {\bibinfo  {journal}
  {Sov. J. Nucl. Phys.}\ }\textbf {\bibinfo {volume} {31}},\ \bibinfo {pages}
  {260} (\bibinfo {year} {1980})},\ \bibinfo {note} {[Yad.
  Fiz. 31, 497 (1980)]}\BibitemShut {NoStop}%
\bibitem [{\citenamefont {{Leinson}}(2014)}]{Leinson2014JCAP}%
  \BibitemOpen
  \bibfield  {author} {\bibinfo {author} {\bibfnamefont {L.~B.}\ \bibnamefont
  {{Leinson}}},\ }\href {\doibase 10.1088/1475-7516/2014/08/031} {\bibfield
  {journal} {\bibinfo  {journal} {\jcap}\ }\textbf {\bibinfo {volume} {08}},\
   (\bibinfo {year} {2014}), 031}.\ \Eprint
  {http://arxiv.org/abs/1405.6873} {arXiv:1405.6873 [hep-ph]} \BibitemShut
  {NoStop}%
\bibitem [{\citenamefont {{Hamaguchi}}\ \emph {et~al.}(2018)\citenamefont
  {{Hamaguchi}}, \citenamefont {{Nagata}}, \citenamefont {{Yanagi}},\ and\
  \citenamefont {{Zheng}}}]{Hamaguchi2018}%
  \BibitemOpen
  \bibfield  {author} {\bibinfo {author} {\bibfnamefont {K.}~\bibnamefont
  {{Hamaguchi}}}, \bibinfo {author} {\bibfnamefont {N.}~\bibnamefont
  {{Nagata}}}, \bibinfo {author} {\bibfnamefont {K.}~\bibnamefont {{Yanagi}}},
  \ and\ \bibinfo {author} {\bibfnamefont {J.}~\bibnamefont {{Zheng}}},\ }\href
  {\doibase 10.1103/PhysRevD.98.103015} {\bibfield  {journal} {\bibinfo
  {journal} {\prd}\ }\textbf {\bibinfo {volume} {98}},\ \bibinfo {eid} {103015}
  (\bibinfo {year} {2018})},\ \Eprint {http://arxiv.org/abs/1806.07151}
  {arXiv:1806.07151 [hep-ph]} \BibitemShut {NoStop}%
\bibitem [{\citenamefont {{Beznogov}}\ \emph {et~al.}()\citenamefont
  {{Beznogov}}, \citenamefont {{Rrapaj}}, \citenamefont {{Page}},\ and\
  \citenamefont {{Reddy}}}]{Beznogov2018}%
  \BibitemOpen
  \bibfield  {author} {\bibinfo {author} {\bibfnamefont {M.~V.}\ \bibnamefont
  {{Beznogov}}}, \bibinfo {author} {\bibfnamefont {E.}~\bibnamefont
  {{Rrapaj}}}, \bibinfo {author} {\bibfnamefont {D.}~\bibnamefont {{Page}}}, \
  and\ \bibinfo {author} {\bibfnamefont {S.}~\bibnamefont {{Reddy}}},\
  }\href@noop {} {\bibinfo  {journal} {\prc}\ 
  \textbf {\bibinfo {volume} {98}},\
  \bibinfo {eid} {035802} (\bibinfo {year} {2018})}
\BibitemShut {NoStop}%
\bibitem [{\citenamefont {{Iwamoto}}(1984)}]{Iwamoto1984}%
  \BibitemOpen
\bibfield  {journal} {  }\bibfield  {author} {\bibinfo {author} {\bibfnamefont
  {N.}~\bibnamefont {{Iwamoto}}},\ }\href {\doibase
  10.1103/PhysRevLett.53.1198} {\bibfield  {journal} {\bibinfo  {journal}
  {\prl}\ }\textbf {\bibinfo {volume} {53}},\ \bibinfo {pages} {1198} (\bibinfo
  {year} {1984})}\BibitemShut {NoStop}%
\bibitem [{\citenamefont {{Nakagawa}}\ \emph {et~al.}(1987)\citenamefont
  {{Nakagawa}}, \citenamefont {{Kohyama}},\ and\ \citenamefont
  {{Itoh}}}]{Nakagawa1987}%
  \BibitemOpen
  \bibfield  {author} {\bibinfo {author} {\bibfnamefont {M.}~\bibnamefont
  {{Nakagawa}}}, \bibinfo {author} {\bibfnamefont {Y.}~\bibnamefont
  {{Kohyama}}}, \ and\ \bibinfo {author} {\bibfnamefont {N.}~\bibnamefont
  {{Itoh}}},\ }\href {\doibase 10.1086/165724} {\bibfield  {journal} {\bibinfo
  {journal} {\apj}\ }\textbf {\bibinfo {volume} {322}},\ \bibinfo {pages} {291}
  (\bibinfo {year} {1987})}\BibitemShut {NoStop}%
\bibitem [{\citenamefont {{Nakagawa}}\ \emph {et~al.}(1988)\citenamefont
  {{Nakagawa}}, \citenamefont {{Adachi}}, \citenamefont {{Kohyama}},\ and\
  \citenamefont {{Itoh}}}]{Nakagawa1988}%
  \BibitemOpen
  \bibfield  {author} {\bibinfo {author} {\bibfnamefont {M.}~\bibnamefont
  {{Nakagawa}}}, \bibinfo {author} {\bibfnamefont {T.}~\bibnamefont
  {{Adachi}}}, \bibinfo {author} {\bibfnamefont {Y.}~\bibnamefont {{Kohyama}}},
  \ and\ \bibinfo {author} {\bibfnamefont {N.}~\bibnamefont {{Itoh}}},\ }\href
  {\doibase 10.1086/166085} {\bibfield  {journal} {\bibinfo  {journal} {\apj}\
  }\textbf {\bibinfo {volume} {326}},\ \bibinfo {pages} {241} (\bibinfo {year}
  {1988})}\BibitemShut {NoStop}%
\bibitem [{\citenamefont {{Umeda}}\ \emph {et~al.}(1998)\citenamefont
  {{Umeda}}, \citenamefont {{Iwamoto}}, \citenamefont {{Tsuruta}},
  \citenamefont {{Qin}},\ and\ \citenamefont {{Nomoto}}}]{Umeda1998}%
  \BibitemOpen
  \bibfield  {author} {\bibinfo {author} {\bibfnamefont {H.}~\bibnamefont
  {{Umeda}}}, \bibinfo {author} {\bibfnamefont {N.}~\bibnamefont {{Iwamoto}}},
  \bibinfo {author} {\bibfnamefont {S.}~\bibnamefont {{Tsuruta}}}, \bibinfo
  {author} {\bibfnamefont {L.}~\bibnamefont {{Qin}}}, \ and\ \bibinfo {author}
  {\bibfnamefont {K.}~\bibnamefont {{Nomoto}}},\ }in\ \href@noop {} {\emph
  {\bibinfo {booktitle} {Neutron Stars and Pulsars: Thirty Years after the
  Discovery}}},\ \bibinfo {editor} {edited by\ \bibinfo {editor} {\bibfnamefont
  {N.}~\bibnamefont {{Shibazaki} }}}\ (\bibinfo {year} {Universal
Academy Press, Tokyo, 1998})\ p.\ \bibinfo
  {pages} {213}\ \BibitemShut {NoStop}%
\bibitem [{\citenamefont {{Altherr}}\ \emph {et~al.}(1994)\citenamefont
  {{Altherr}}, \citenamefont {{Petitgirard}},\ and\ \citenamefont {{del
  R{\'{\i}}o\^{}Gaztelurrutia}}}]{1994APh.....2..175A}%
  \BibitemOpen
  \bibfield  {author} {\bibinfo {author} {\bibfnamefont {T.}~\bibnamefont
  {{Altherr}}}, \bibinfo {author} {\bibfnamefont {E.}~\bibnamefont
  {{Petitgirard}}}, \ and\ \bibinfo {author} {\bibfnamefont {T.}~\bibnamefont
  {{del R{\'{\i}}o\^{}Gaztelurrutia}}},\ }\href {\doibase
  10.1016/0927-6505(94)90040-X} {\bibfield  {journal} {\bibinfo  {journal}
  {Astropart. Phys.}\ }\textbf {\bibinfo {volume} {2}},\ \bibinfo {pages} {175}
  (\bibinfo {year} {1994})},\ \Eprint {http://arxiv.org/abs/hep-ph/9310304}
  {hep-ph/9310304} \BibitemShut {NoStop}%
\bibitem [{\citenamefont {{Raffelt}}\ and\ \citenamefont
  {{Weiss}}(1995)}]{1995PhRvD..51.1495R}%
  \BibitemOpen
  \bibfield  {author} {\bibinfo {author} {\bibfnamefont {G.}~\bibnamefont
  {{Raffelt}}}\ and\ \bibinfo {author} {\bibfnamefont {A.}~\bibnamefont
  {{Weiss}}},\ }\href {\doibase 10.1103/PhysRevD.51.1495} {\bibfield  {journal}
  {\bibinfo  {journal} {\prd}\ }\textbf {\bibinfo {volume} {51}},\ \bibinfo
  {pages} {1495} (\bibinfo {year} {1995})},\ \Eprint
  {http://arxiv.org/abs/hep-ph/9410205} {hep-ph/9410205} \BibitemShut {NoStop}%
\bibitem [{\citenamefont {{C{\'o}rsico}}\ \emph {et~al.}(2001)\citenamefont
  {{C{\'o}rsico}}, \citenamefont {{Benvenuto}}, \citenamefont {{Althaus}},
  \citenamefont {{Isern}},\ and\ \citenamefont
  {{Garc{\'{\i}}a-Berro}}}]{2001NewA....6..197C}%
  \BibitemOpen
  \bibfield  {author} {\bibinfo {author} {\bibfnamefont {A.~H.}\ \bibnamefont
  {{C{\'o}rsico}}}, \bibinfo {author} {\bibfnamefont {O.~G.}\ \bibnamefont
  {{Benvenuto}}}, \bibinfo {author} {\bibfnamefont {L.~G.}\ \bibnamefont
  {{Althaus}}}, \bibinfo {author} {\bibfnamefont {J.}~\bibnamefont {{Isern}}},
  \ and\ \bibinfo {author} {\bibfnamefont {E.}~\bibnamefont
  {{Garc{\'{\i}}a-Berro}}},\ }\href {\doibase 10.1016/S1384-1076(01)00055-0}
  {\bibfield  {journal} {\bibinfo  {journal} {\na}\ }\textbf {\bibinfo {volume}
  {6}},\ \bibinfo {pages} {197} (\bibinfo {year} {2001})},\ \Eprint
  {http://arxiv.org/abs/astro-ph/0104103} {astro-ph/0104103} \BibitemShut
  {NoStop}%
\bibitem [{\citenamefont {{Miller Bertolami}}\ \emph
  {et~al.}(2014)\citenamefont {{Miller Bertolami}}, \citenamefont {{Melendez}},
  \citenamefont {{Althaus}},\ and\ \citenamefont
  {{Isern}}}]{2014JCAP...10..069M}%
  \BibitemOpen
  \bibfield  {author} {\bibinfo {author} {\bibfnamefont {M.~M.}\ \bibnamefont
  {{Miller Bertolami}}}, \bibinfo {author} {\bibfnamefont {B.~E.}\ \bibnamefont
  {{Melendez}}}, \bibinfo {author} {\bibfnamefont {L.~G.}\ \bibnamefont
  {{Althaus}}}, \ and\ \bibinfo {author} {\bibfnamefont {J.}~\bibnamefont
  {{Isern}}},\ }\href {\doibase 10.1088/1475-7516/2014/10/069} {\bibfield
  {journal} {\bibinfo  {journal} {\jcap}\ }\textbf {\bibinfo {volume} {10}},\
  \bibinfo {eid} {069} (\bibinfo {year} {2014})},\ \Eprint
  {http://arxiv.org/abs/1406.7712} {arXiv:1406.7712 [hep-ph]} \BibitemShut
  {NoStop}%
\bibitem [{\citenamefont {{Brinkmann}}\ and\ \citenamefont
  {{Turner}}(1988)}]{1988PhRvD..38.2338B}%
  \BibitemOpen
  \bibfield  {author} {\bibinfo {author} {\bibfnamefont {R.~P.}\ \bibnamefont
  {{Brinkmann}}}\ and\ \bibinfo {author} {\bibfnamefont {M.~S.}\ \bibnamefont
  {{Turner}}},\ }\href {\doibase 10.1103/PhysRevD.38.2338} {\bibfield
  {journal} {\bibinfo  {journal} {\prd}\ }\textbf {\bibinfo {volume} {38}},\
  \bibinfo {pages} {2338} (\bibinfo {year} {1988})}\BibitemShut {NoStop}%
\bibitem [{\citenamefont {{Burrows}}\ \emph {et~al.}(1989)\citenamefont
  {{Burrows}}, \citenamefont {{Turner}},\ and\ \citenamefont
  {{Brinkmann}}}]{1989PhRvD..39.1020B}%
  \BibitemOpen
  \bibfield  {author} {\bibinfo {author} {\bibfnamefont {A.}~\bibnamefont
  {{Burrows}}}, \bibinfo {author} {\bibfnamefont {M.~S.}\ \bibnamefont
  {{Turner}}}, \ and\ \bibinfo {author} {\bibfnamefont {R.~P.}\ \bibnamefont
  {{Brinkmann}}},\ }\href {\doibase 10.1103/PhysRevD.39.1020} {\bibfield
  {journal} {\bibinfo  {journal} {\prd}\ }\textbf {\bibinfo {volume} {39}},\
  \bibinfo {pages} {1020} (\bibinfo {year} {1989})}\BibitemShut {NoStop}%
\bibitem [{\citenamefont {{Burrows}}\ \emph {et~al.}(1990)\citenamefont
  {{Burrows}}, \citenamefont {{Ressell}},\ and\ \citenamefont
  {{Turner}}}]{1990PhRvD..42.3297B}%
  \BibitemOpen
  \bibfield  {author} {\bibinfo {author} {\bibfnamefont {A.}~\bibnamefont
  {{Burrows}}}, \bibinfo {author} {\bibfnamefont {M.~T.}\ \bibnamefont
  {{Ressell}}}, \ and\ \bibinfo {author} {\bibfnamefont {M.~S.}\ \bibnamefont
  {{Turner}}},\ }\href {\doibase 10.1103/PhysRevD.42.3297} {\bibfield
  {journal} {\bibinfo  {journal} {\prd}\ }\textbf {\bibinfo {volume} {42}},\
  \bibinfo {pages} {3297} (\bibinfo {year} {1990})}\BibitemShut {NoStop}%
\bibitem [{\citenamefont {{Janka}}\ \emph {et~al.}(1996)\citenamefont
  {{Janka}}, \citenamefont {{Keil}}, \citenamefont {{Raffelt}},\ and\
  \citenamefont {{Seckel}}}]{1996PhRvL..76.2621J}%
  \BibitemOpen
  \bibfield  {author} {\bibinfo {author} {\bibfnamefont {H.-T.}\ \bibnamefont
  {{Janka}}}, \bibinfo {author} {\bibfnamefont {W.}~\bibnamefont {{Keil}}},
  \bibinfo {author} {\bibfnamefont {G.}~\bibnamefont {{Raffelt}}}, \ and\
  \bibinfo {author} {\bibfnamefont {D.}~\bibnamefont {{Seckel}}},\ }\href
  {\doibase 10.1103/PhysRevLett.76.2621} {\bibfield  {journal} {\bibinfo
  {journal} {\prl}\ }\textbf {\bibinfo {volume} {76}},\ \bibinfo {pages} {2621}
  (\bibinfo {year} {1996})},\ \Eprint {http://arxiv.org/abs/astro-ph/9507023}
  {astro-ph/9507023} \BibitemShut {NoStop}%
\bibitem [{\citenamefont {{Hanhart}}\ \emph {et~al.}(2001)\citenamefont
  {{Hanhart}}, \citenamefont {{Phillips}},\ and\ \citenamefont
  {{Reddy}}}]{2001PhLB..499....9H}%
  \BibitemOpen
  \bibfield  {author} {\bibinfo {author} {\bibfnamefont {C.}~\bibnamefont
  {{Hanhart}}}, \bibinfo {author} {\bibfnamefont {D.~R.}\ \bibnamefont
  {{Phillips}}}, \ and\ \bibinfo {author} {\bibfnamefont {S.}~\bibnamefont
  {{Reddy}}},\ }\href {\doibase 10.1016/S0370-2693(00)01382-4} {\bibfield
  {journal} {\bibinfo  {journal} {Phys. Lett. B}\ }\textbf {\bibinfo {volume}
  {499}},\ \bibinfo {pages} {9} (\bibinfo {year} {2001})} \BibitemShut
  {NoStop}%
\bibitem [{\citenamefont {{Fischer}}\ \emph {et~al.}(2016)\citenamefont
  {{Fischer}}, \citenamefont {{Chakraborty}}, \citenamefont {{Giannotti}},
  \citenamefont {{Mirizzi}}, \citenamefont {{Payez}},\ and\ \citenamefont
  {{Ringwald}}}]{2016PhRvD..94h5012F}%
  \BibitemOpen
  \bibfield  {author} {\bibinfo {author} {\bibfnamefont {T.}~\bibnamefont
  {{Fischer}}}, \bibinfo {author} {\bibfnamefont {S.}~\bibnamefont
  {{Chakraborty}}}, \bibinfo {author} {\bibfnamefont {M.}~\bibnamefont
  {{Giannotti}}}, \bibinfo {author} {\bibfnamefont {A.}~\bibnamefont
  {{Mirizzi}}}, \bibinfo {author} {\bibfnamefont {A.}~\bibnamefont {{Payez}}},
  \ and\ \bibinfo {author} {\bibfnamefont {A.}~\bibnamefont {{Ringwald}}},\
  }\href {\doibase 10.1103/PhysRevD.94.085012} {\bibfield  {journal} {\bibinfo
  {journal} {\prd}\ }\textbf {\bibinfo {volume} {94}},\ \bibinfo {eid} {085012}
  (\bibinfo {year} {2016})},\ \Eprint {http://arxiv.org/abs/1605.08780}
  {arXiv:1605.08780 [astro-ph.HE]} \BibitemShut {NoStop}%
\bibitem [{\citenamefont {{Keller}}\ and\ \citenamefont
  {{Sedrakian}}(2013)}]{2013NuPhA.897...62K}%
  \BibitemOpen
  \bibfield  {author} {\bibinfo {author} {\bibfnamefont {J.}~\bibnamefont
  {{Keller}}}\ and\ \bibinfo {author} {\bibfnamefont {A.}~\bibnamefont
  {{Sedrakian}}},\ }\href {\doibase 10.1016/j.nuclphysa.2012.11.004} {\bibfield
   {journal} {\bibinfo  {journal} {Nucl. Phys.}\ }\textbf {\bibinfo {volume}
  {A897}},\ \bibinfo {pages} {62} (\bibinfo {year} {2013})},\ \Eprint
  {http://arxiv.org/abs/1205.6940} {arXiv:1205.6940 [astro-ph.CO]} \BibitemShut
  {NoStop}%
\bibitem [{\citenamefont {{Yakovlev}}\ \emph {et~al.}(1999)\citenamefont
  {{Yakovlev}}, \citenamefont {{Kaminker}},\ and\ \citenamefont
  {{Levenfish}}}]{1999A&A...343..650Y}%
  \BibitemOpen
  \bibfield  {author} {\bibinfo {author} {\bibfnamefont {D.~G.}\ \bibnamefont
  {{Yakovlev}}}, \bibinfo {author} {\bibfnamefont {A.~D.}\ \bibnamefont
  {{Kaminker}}}, \ and\ \bibinfo {author} {\bibfnamefont {K.~P.}\ \bibnamefont
  {{Levenfish}}},\ }\href@noop {} {\bibfield  {journal} {\bibinfo  {journal}
  {\aap}\ }\textbf {\bibinfo {volume} {343}},\ \bibinfo {pages} {650} (\bibinfo
  {year} {1999})},\ \Eprint {http://arxiv.org/abs/astro-ph/9812366}
  {astro-ph/9812366} \BibitemShut {NoStop}%
\bibitem [{\citenamefont {{Leinson}}\ and\ \citenamefont
  {{P{\'e}rez}}(2006)}]{2006PhLB..638..114L}%
  \BibitemOpen
  \bibfield  {author} {\bibinfo {author} {\bibfnamefont {L.~B.}\ \bibnamefont
  {{Leinson}}}\ and\ \bibinfo {author} {\bibfnamefont {A.}~\bibnamefont
  {{P{\'e}rez}}},\ }\href {\doibase 10.1016/j.physletb.2006.05.036} {\bibfield
  {journal} {\bibinfo  {journal} {Phys. Lett. B}\ }\textbf {\bibinfo {volume}
  {638}},\ \bibinfo {pages} {114} (\bibinfo {year} {2006})},\ \Eprint
  {http://arxiv.org/abs/astro-ph/0606651} {astro-ph/0606651} \BibitemShut
  {NoStop}%
\bibitem [{\citenamefont {{Sedrakian}}\ \emph {et~al.}(2007)\citenamefont
  {{Sedrakian}}, \citenamefont {{M{\"u}ther}},\ and\ \citenamefont
  {{Schuck}}}]{2007PhRvC..76e5805S}%
  \BibitemOpen
  \bibfield  {author} {\bibinfo {author} {\bibfnamefont {A.}~\bibnamefont
  {{Sedrakian}}}, \bibinfo {author} {\bibfnamefont {H.}~\bibnamefont
  {{M{\"u}ther}}}, \ and\ \bibinfo {author} {\bibfnamefont {P.}~\bibnamefont
  {{Schuck}}},\ }\href {\doibase 10.1103/PhysRevC.76.055805} {\bibfield
  {journal} {\bibinfo  {journal} {\prc}\ }\textbf {\bibinfo {volume} {76}},\
  \bibinfo {eid} {055805} (\bibinfo {year} {2007})},\ \Eprint
  {http://arxiv.org/abs/arXiv:astro-ph/0611676} {arXiv:astro-ph/0611676}
  \BibitemShut {NoStop}%
\bibitem [{\citenamefont {{Kolomeitsev}}\ and\ \citenamefont
  {{Voskresensky}}(2008)}]{2008PhRvC..77f5808K}%
  \BibitemOpen
  \bibfield  {author} {\bibinfo {author} {\bibfnamefont {E.~E.}\ \bibnamefont
  {{Kolomeitsev}}}\ and\ \bibinfo {author} {\bibfnamefont {D.~N.}\ \bibnamefont
  {{Voskresensky}}},\ }\href {\doibase 10.1103/PhysRevC.77.065808} {\bibfield
  {journal} {\bibinfo  {journal} {\prc}\ }\textbf {\bibinfo {volume} {77}},\
  \bibinfo {eid} {065808} (\bibinfo {year} {2008})},\ \Eprint
  {http://arxiv.org/abs/0802.1404} {arXiv:0802.1404 [nucl-th]} \BibitemShut
  {NoStop}%
\bibitem [{\citenamefont {{Sedrakian}}(2012)}]{2012PhRvC..86b5803S}%
  \BibitemOpen
  \bibfield  {author} {\bibinfo {author} {\bibfnamefont {A.}~\bibnamefont
  {{Sedrakian}}},\ }\href {\doibase 10.1103/PhysRevC.86.025803} {\bibfield
  {journal} {\bibinfo  {journal} {\prc}\ }\textbf {\bibinfo {volume} {86}},\
  \bibinfo {eid} {025803} (\bibinfo {year} {2012})},\ \Eprint
  {http://arxiv.org/abs/1201.1394} {arXiv:1201.1394 [astro-ph.SR]} \BibitemShut
  {NoStop}%
\bibitem [{\citenamefont {{Harutyunyan}}\ and\ \citenamefont
  {{Sedrakian}}(2016)}]{Harutyunyan2016}%
  \BibitemOpen
  \bibfield  {author} {\bibinfo {author} {\bibfnamefont {A.}~\bibnamefont
  {{Harutyunyan}}}\ and\ \bibinfo {author} {\bibfnamefont {A.}~\bibnamefont
  {{Sedrakian}}},\ }\href {\doibase 10.1103/PhysRevC.94.025805} {\bibfield
  {journal} {\bibinfo  {journal} {\prc}\ }\textbf {\bibinfo {volume} {94}},\
  \bibinfo {eid} {025805} (\bibinfo {year} {2016})}\BibitemShut {NoStop}%
\bibitem [{\citenamefont {{Vigan{\`o}}}\ \emph {et~al.}(2013)\citenamefont
  {{Vigan{\`o}}}, \citenamefont {{Rea}}, \citenamefont {{Pons}}, \citenamefont
  {{Perna}}, \citenamefont {{Aguilera}},\ and\ \citenamefont
  {{Miralles}}}]{2013MNRAS.434..123V}%
  \BibitemOpen
  \bibfield  {author} {\bibinfo {author} {\bibfnamefont {D.}~\bibnamefont
  {{Vigan{\`o}}}}, \bibinfo {author} {\bibfnamefont {N.}~\bibnamefont {{Rea}}},
  \bibinfo {author} {\bibfnamefont {J.~A.}\ \bibnamefont {{Pons}}}, \bibinfo
  {author} {\bibfnamefont {R.}~\bibnamefont {{Perna}}}, \bibinfo {author}
  {\bibfnamefont {D.~N.}\ \bibnamefont {{Aguilera}}}, \ and\ \bibinfo {author}
  {\bibfnamefont {J.~A.}\ \bibnamefont {{Miralles}}},\ }\href {\doibase
  10.1093/mnras/stt1008} {\bibfield  {journal} {\bibinfo  {journal} {\mnras}\
  }\textbf {\bibinfo {volume} {434}},\ \bibinfo {pages} {123} (\bibinfo {year}
  {2013})},\ \Eprint {http://arxiv.org/abs/1306.2156} {arXiv:1306.2156
  [astro-ph.SR]} \BibitemShut {NoStop}%
\bibitem [{\citenamefont {{Page}}\ \emph {et~al.}(2009)\citenamefont {{Page}},
  \citenamefont {{Lattimer}}, \citenamefont {{Prakash}},\ and\ \citenamefont
  {{Steiner}}}]{2009ApJ...707.1131P}%
  \BibitemOpen
  \bibfield  {author} {\bibinfo {author} {\bibfnamefont {D.}~\bibnamefont
  {{Page}}}, \bibinfo {author} {\bibfnamefont {J.~M.}\ \bibnamefont
  {{Lattimer}}}, \bibinfo {author} {\bibfnamefont {M.}~\bibnamefont
  {{Prakash}}}, \ and\ \bibinfo {author} {\bibfnamefont {A.~W.}\ \bibnamefont
  {{Steiner}}},\ }\href {\doibase 10.1088/0004-637X/707/2/1131} {\bibfield
  {journal} {\bibinfo  {journal} {\apj}\ }\textbf {\bibinfo {volume} {707}},\
  \bibinfo {pages} {1131} (\bibinfo {year} {2009})},\ \Eprint
  {http://arxiv.org/abs/0906.1621} {arXiv:0906.1621 [astro-ph.SR]} \BibitemShut
  {NoStop}%
\bibitem [{\citenamefont {{Takatsuka}}(1973)}]{Takatsuka1973}%
  \BibitemOpen
  \bibfield  {author} {\bibinfo {author} {\bibfnamefont {T.}~\bibnamefont
  {{Takatsuka}}},\ }\href {\doibase 10.1143/PTP.50.1754} {\bibfield  {journal}
  {\bibinfo  {journal} {Prog.  Theor. Phys.}\ }\textbf {\bibinfo
  {volume} {50}},\ \bibinfo {pages} {1754} (\bibinfo {year}
  {1973})}\BibitemShut {NoStop}%
\bibitem [{\citenamefont {{Wambach}}\ \emph
  {et~al.}(1993{\natexlab{a}})\citenamefont {{Wambach}}, \citenamefont
  {{Ainsworth}},\ and\ \citenamefont {{Pines}}}]{WAP1993}%
  \BibitemOpen
  \bibfield  {author} {\bibinfo {author} {\bibfnamefont {J.}~\bibnamefont
  {{Wambach}}}, \bibinfo {author} {\bibfnamefont {T.~L.}\ \bibnamefont
  {{Ainsworth}}}, \ and\ \bibinfo {author} {\bibfnamefont {D.}~\bibnamefont
  {{Pines}}},\ }\href {\doibase 10.1016/0375-9474(93)90317-Q} {\bibfield
  {journal} {\bibinfo  {journal} {Nucl. Phys.}\ }\textbf {\bibinfo {volume}
  {A555}},\ \bibinfo {pages} {128} (\bibinfo {year}
  {1993}{\natexlab{a}})}\BibitemShut {NoStop}%
\bibitem [{\citenamefont {{Fan}}\ \emph {et~al.}(2017)\citenamefont {{Fan}},
  \citenamefont {{Krotscheck}},\ and\ \citenamefont
  {{Clark}}}]{FanKrotscheck2017}%
  \BibitemOpen
  \bibfield  {author} {\bibinfo {author} {\bibfnamefont {H.-H.}\ \bibnamefont
  {{Fan}}}, \bibinfo {author} {\bibfnamefont {E.}~\bibnamefont {{Krotscheck}}},
  \ and\ \bibinfo {author} {\bibfnamefont {J.~W.}\ \bibnamefont {{Clark}}},\
  }\href {\doibase 10.1007/s10909-017-1813-z} {\bibfield  {journal} {\bibinfo
  {journal} {J. Low Temp. Phys.}\ }\textbf {\bibinfo {volume} {189}},\
  \bibinfo {pages} {470} (\bibinfo {year} {2017})},\ \Eprint
  {http://arxiv.org/abs/1707.07268} {arXiv:1707.07268 [nucl-th]} \BibitemShut
  {NoStop}%
\bibitem [{\citenamefont {{Page}}\ \emph {et~al.}(2004)\citenamefont {{Page}},
  \citenamefont {{Lattimer}}, \citenamefont {{Prakash}},\ and\ \citenamefont
  {{Steiner}}}]{2004ApJS..155..623P}%
  \BibitemOpen
  \bibfield  {author} {\bibinfo {author} {\bibfnamefont {D.}~\bibnamefont
  {{Page}}}, \bibinfo {author} {\bibfnamefont {J.~M.}\ \bibnamefont
  {{Lattimer}}}, \bibinfo {author} {\bibfnamefont {M.}~\bibnamefont
  {{Prakash}}}, \ and\ \bibinfo {author} {\bibfnamefont {A.~W.}\ \bibnamefont
  {{Steiner}}},\ }\href {\doibase 10.1086/424844} {\bibfield  {journal}
  {\bibinfo  {journal} {\apjs}\ }\textbf {\bibinfo {volume} {155}},\ \bibinfo
  {pages} {623} (\bibinfo {year} {2004})},\ \Eprint
  {http://arxiv.org/abs/arXiv:astro-ph/0403657} {arXiv:astro-ph/0403657}
  \BibitemShut {NoStop}%
\bibitem [{\citenamefont {{Chen}}\ \emph {et~al.}(1993)\citenamefont {{Chen}},
  \citenamefont {{Clark}}, \citenamefont {{Dav{\'e}}},\ and\ \citenamefont
  {{Khodel}}}]{CCDK1993}%
  \BibitemOpen
  \bibfield  {author} {\bibinfo {author} {\bibfnamefont {J.~M.~C.}\
  \bibnamefont {{Chen}}}, \bibinfo {author} {\bibfnamefont {J.~W.}\
  \bibnamefont {{Clark}}}, \bibinfo {author} {\bibfnamefont {R.~D.}\
  \bibnamefont {{Dav{\'e}}}}, \ and\ \bibinfo {author} {\bibfnamefont {V.~V.}\
  \bibnamefont {{Khodel}}},\ }\href {\doibase 10.1016/0375-9474(93)90314-N}
  {\bibfield  {journal} {\bibinfo  {journal} {Nucl. Phys.}\ }\textbf
  {\bibinfo {volume} {A555}},\ \bibinfo {pages} {59} (\bibinfo {year}
  {1993})}\BibitemShut {NoStop}%
\bibitem [{\citenamefont {{Gotthelf}}\ \emph {et~al.}(2013)\citenamefont
  {{Gotthelf}}, \citenamefont {{Halpern}},\ and\ \citenamefont
  {{Alford}}}]{2013ApJ...765...58G}%
  \BibitemOpen
  \bibfield  {author} {\bibinfo {author} {\bibfnamefont {E.~V.}\ \bibnamefont
  {{Gotthelf}}}, \bibinfo {author} {\bibfnamefont {J.~P.}\ \bibnamefont
  {{Halpern}}}, \ and\ \bibinfo {author} {\bibfnamefont {J.}~\bibnamefont
  {{Alford}}},\ }\href {\doibase 10.1088/0004-637X/765/1/58} {\bibfield
  {journal} {\bibinfo  {journal} {\apj}\ }\textbf {\bibinfo {volume} {765}},\
  \bibinfo {eid} {58} (\bibinfo {year} {2013})},\ \Eprint
  {http://arxiv.org/abs/1301.2717} {arXiv:1301.2717 [astro-ph.HE]} \BibitemShut
  {NoStop}%
\bibitem [{\citenamefont {{Elshamouty}}\ \emph {et~al.}(2013)\citenamefont
  {{Elshamouty}}, \citenamefont {{Heinke}}, \citenamefont {{Sivakoff}},
  \citenamefont {{Ho}}, \citenamefont {{Shternin}}, \citenamefont {{Yakovlev}},
  \citenamefont {{Patnaude}},\ and\ \citenamefont
  {{David}}}]{2013ApJ...777...22E}%
  \BibitemOpen
  \bibfield  {author} {\bibinfo {author} {\bibfnamefont {K.~G.}\ \bibnamefont
  {{Elshamouty}}}, \bibinfo {author} {\bibfnamefont {C.~O.}\ \bibnamefont
  {{Heinke}}}, \bibinfo {author} {\bibfnamefont {G.~R.}\ \bibnamefont
  {{Sivakoff}}}, \bibinfo {author} {\bibfnamefont {W.~C.~G.}\ \bibnamefont
  {{Ho}}}, \bibinfo {author} {\bibfnamefont {P.~S.}\ \bibnamefont
  {{Shternin}}}, \bibinfo {author} {\bibfnamefont {D.~G.}\ \bibnamefont
  {{Yakovlev}}}, \bibinfo {author} {\bibfnamefont {D.~J.}\ \bibnamefont
  {{Patnaude}}}, \ and\ \bibinfo {author} {\bibfnamefont {L.}~\bibnamefont
  {{David}}},\ }\href {\doibase 10.1088/0004-637X/777/1/22} {\bibfield
  {journal} {\bibinfo  {journal} {\apj}\ }\textbf {\bibinfo {volume} {777}},\
  \bibinfo {eid} {22} (\bibinfo {year} {2013})},\ \Eprint
  {http://arxiv.org/abs/1306.3387} {arXiv:1306.3387 [astro-ph.HE]} \BibitemShut
  {NoStop}%
\bibitem [{\citenamefont {{De Luca}}\ \emph {et~al.}(2005)\citenamefont {{De
  Luca}}, \citenamefont {{Caraveo}}, \citenamefont {{Mereghetti}},
  \citenamefont {{Negroni}},\ and\ \citenamefont
  {{Bignami}}}]{2005ApJ...623.1051D}%
  \BibitemOpen
  \bibfield  {author} {\bibinfo {author} {\bibfnamefont {A.}~\bibnamefont {{De
  Luca}}}, \bibinfo {author} {\bibfnamefont {P.~A.}\ \bibnamefont {{Caraveo}}},
  \bibinfo {author} {\bibfnamefont {S.}~\bibnamefont {{Mereghetti}}}, \bibinfo
  {author} {\bibfnamefont {M.}~\bibnamefont {{Negroni}}}, \ and\ \bibinfo
  {author} {\bibfnamefont {G.~F.}\ \bibnamefont {{Bignami}}},\ }\href {\doibase
  10.1086/428567} {\bibfield  {journal} {\bibinfo  {journal} {\apj}\ }\textbf
  {\bibinfo {volume} {623}},\ \bibinfo {pages} {1051} (\bibinfo {year}
  {2005})},\ \Eprint {http://arxiv.org/abs/astro-ph/0412662} {astro-ph/0412662}
  \BibitemShut {NoStop}%
\bibitem [{\citenamefont {{Posselt}}\ and\ \citenamefont
  {{Pavlov}}(2018)}]{Posselt2018}%
  \BibitemOpen
  \bibfield  {author} {\bibinfo {author} {\bibfnamefont {B.}~\bibnamefont
  {{Posselt}}}\ and\ \bibinfo {author} {\bibfnamefont {G.~G.}\ \bibnamefont
  {{Pavlov}}},\ }\href {\doibase 10.3847/1538-4357/aad7fc} {\bibfield
  {journal} {\bibinfo  {journal} {\apj}\ }\textbf {\bibinfo {volume} {864}},\
  \bibinfo {eid} {135} (\bibinfo {year} {2018})},\ \Eprint
  {http://arxiv.org/abs/1808.00531} {arXiv:1808.00531 [astro-ph.HE]}
  \BibitemShut {NoStop}%
\bibitem [{\citenamefont {{Blaschke}}\ \emph {et~al.}(2001)\citenamefont
  {{Blaschke}}, \citenamefont {{Grigorian}},\ and\ \citenamefont
  {{Voskresensky}}}]{2001A&A...368..561B}%
  \BibitemOpen
  \bibfield  {author} {\bibinfo {author} {\bibfnamefont {D.}~\bibnamefont
  {{Blaschke}}}, \bibinfo {author} {\bibfnamefont {H.}~\bibnamefont
  {{Grigorian}}}, \ and\ \bibinfo {author} {\bibfnamefont {D.~N.}\ \bibnamefont
  {{Voskresensky}}},\ }\href {\doibase 10.1051/0004-6361:20010005} {\bibfield
  {journal} {\bibinfo  {journal} {\aap}\ }\textbf {\bibinfo {volume} {368}},\
  \bibinfo {pages} {561} (\bibinfo {year} {2001})},\ \Eprint
  {http://arxiv.org/abs/astro-ph/0009120} {astro-ph/0009120} \BibitemShut
  {NoStop}%
\bibitem [{\citenamefont {{Hess}}\ and\ \citenamefont
  {{Sedrakian}}(2011)}]{2011PhRvD..84f3015H}%
  \BibitemOpen
  \bibfield  {author} {\bibinfo {author} {\bibfnamefont {D.}~\bibnamefont
  {{Hess}}}\ and\ \bibinfo {author} {\bibfnamefont {A.}~\bibnamefont
  {{Sedrakian}}},\ }\href {\doibase 10.1103/PhysRevD.84.063015} {\bibfield
  {journal} {\bibinfo  {journal} {\prd}\ }\textbf {\bibinfo {volume} {84}},\
  \bibinfo {eid} {063015} (\bibinfo {year} {2011})},\ \Eprint
  {http://arxiv.org/abs/1104.1706} {arXiv:1104.1706 [astro-ph.HE]} \BibitemShut
  {NoStop}%
\bibitem [{\citenamefont {{Alford}}\ \emph {et~al.}(2005)\citenamefont
  {{Alford}}, \citenamefont {{Jotwani}}, \citenamefont {{Kouvaris}},
  \citenamefont {{Kundu}},\ and\ \citenamefont
  {{Rajagopal}}}]{2005PhRvD..71k4011A}%
  \BibitemOpen
  \bibfield  {author} {\bibinfo {author} {\bibfnamefont {M.}~\bibnamefont
  {{Alford}}}, \bibinfo {author} {\bibfnamefont {P.}~\bibnamefont {{Jotwani}}},
  \bibinfo {author} {\bibfnamefont {C.}~\bibnamefont {{Kouvaris}}}, \bibinfo
  {author} {\bibfnamefont {J.}~\bibnamefont {{Kundu}}}, \ and\ \bibinfo
  {author} {\bibfnamefont {K.}~\bibnamefont {{Rajagopal}}},\ }\href {\doibase
  10.1103/PhysRevD.71.114011} {\bibfield  {journal} {\bibinfo  {journal}
  {\prd}\ }\textbf {\bibinfo {volume} {71}},\ \bibinfo {eid} {114011} (\bibinfo
  {year} {2005})},\ \Eprint {http://arxiv.org/abs/astro-ph/0411560}
  {astro-ph/0411560} \BibitemShut {NoStop}%
\bibitem [{\citenamefont
  {{Sedrakian}}(2016{\natexlab{b}})}]{2016EPJA...52...44S}%
  \BibitemOpen
  \bibfield  {author} {\bibinfo {author} {\bibfnamefont {A.}~\bibnamefont
  {{Sedrakian}}},\ }\href {\doibase 10.1140/epja/i2016-16044-y} {\bibfield
  {journal} {\bibinfo  {journal} {Eur. Phys. J. A}\ }\textbf {\bibinfo {volume} {52}},\
  \bibinfo {eid} {44} (\bibinfo {year} {2016}{\natexlab{b}})},\ \Eprint
  {http://arxiv.org/abs/1509.06986} {arXiv:1509.06986 [astro-ph.HE]}
  \BibitemShut {NoStop}%
\bibitem [{\citenamefont {{Raduta}}\ \emph {et~al.}(2018)\citenamefont
  {{Raduta}}, \citenamefont {{Sedrakian}},\ and\ \citenamefont
  {{Weber}}}]{2018MNRAS.475.4347R}%
  \BibitemOpen
  \bibfield  {author} {\bibinfo {author} {\bibfnamefont {A.~R.}\ \bibnamefont
  {{Raduta}}}, \bibinfo {author} {\bibfnamefont {A.}~\bibnamefont
  {{Sedrakian}}}, \ and\ \bibinfo {author} {\bibfnamefont {F.}~\bibnamefont
  {{Weber}}},\ }\href {\doibase 10.1093/mnras/stx3318} {\bibfield  {journal}
  {\bibinfo  {journal} {\mnras}\ }\textbf {\bibinfo {volume} {475}},\ \bibinfo
  {pages} {4347} (\bibinfo {year} {2018})},\ \Eprint
  {http://arxiv.org/abs/1712.00584} {arXiv:1712.00584 [astro-ph.HE]}
  \BibitemShut {NoStop}%
\bibitem [{\citenamefont {{Grigorian}}\ \emph {et~al.}(2018)\citenamefont
  {{Grigorian}}, \citenamefont {{Voskresensky}},\ and\ \citenamefont
  {{Maslov}}}]{2018arXiv180801819G}%
  \BibitemOpen
  \bibfield  {author} {\bibinfo {author} {\bibfnamefont {H.}~\bibnamefont
  {{Grigorian}}}, \bibinfo {author} {\bibfnamefont {D.~N.}\ \bibnamefont
  {{Voskresensky}}}, \ and\ \bibinfo {author} {\bibfnamefont {K.~A.}\
  \bibnamefont {{Maslov}}},\ }\href@noop {} {\bibfield  {journal} {\bibinfo
  {journal} {Nucl. Phys. {\bf A980}, 105 }\ } (\bibinfo {year} {2018})},\ \Eprint
  {http://arxiv.org/abs/1808.01819} {arXiv:1808.01819 [astro-ph.HE]}
  \BibitemShut {NoStop}%
\bibitem [{\citenamefont {{Negreiros}}\ \emph {et~al.}(2018)\citenamefont
  {{Negreiros}}, \citenamefont {{Tolos}}, \citenamefont {{Centelles}},
  \citenamefont {{Ramos}},\ and\ \citenamefont
  {{Dexheimer}}}]{2018ApJ...863..104N}%
  \BibitemOpen
  \bibfield  {author} {\bibinfo {author} {\bibfnamefont {R.}~\bibnamefont
  {{Negreiros}}}, \bibinfo {author} {\bibfnamefont {L.}~\bibnamefont
  {{Tolos}}}, \bibinfo {author} {\bibfnamefont {M.}~\bibnamefont
  {{Centelles}}}, \bibinfo {author} {\bibfnamefont {A.}~\bibnamefont
  {{Ramos}}}, \ and\ \bibinfo {author} {\bibfnamefont {V.}~\bibnamefont
  {{Dexheimer}}},\ }\href {\doibase 10.3847/1538-4357/aad049} {\bibfield
  {journal} {\bibinfo  {journal} {\apj}\ }\textbf {\bibinfo {volume} {863}},\
  \bibinfo {eid} {104} (\bibinfo {year} {2018})},\ \Eprint
  {http://arxiv.org/abs/1804.00334} {arXiv:1804.00334 [astro-ph.HE]}
  \BibitemShut {NoStop}%
\bibitem [{\citenamefont {Weber}(1999)}]{weber_book}%
  \BibitemOpen
  \bibfield  {author} {\bibinfo {author} {\bibfnamefont {F.}~\bibnamefont
  {Weber}},\ }\href@noop {} {\emph {\bibinfo {title} {Pulsars as Astrophysical
  Laboratories for Nuclear and Particle Physics}}}\ (\bibinfo  {publisher}
  {Institute of Physics},\ \bibinfo {address} {Bristol, UK},\ \bibinfo {year}
  {1999})\BibitemShut {NoStop}%
\bibitem [{\citenamefont {{Sedrakian}}(2007)}]{2007PrPNP..58..168S}%
  \BibitemOpen
  \bibfield  {author} {\bibinfo {author} {\bibfnamefont {A.}~\bibnamefont
  {{Sedrakian}}},\ }\href {\doibase 10.1016/j.ppnp.2006.02.002} {\bibfield
  {journal} {\bibinfo  {journal} {Prog. Part. Nucl. Phys.}\ }\textbf {\bibinfo
  {volume} {58}},\ \bibinfo {pages} {168} (\bibinfo {year} {2007})},\ \Eprint
  {http://arxiv.org/abs/arXiv:nucl-th/0601086} {arXiv:nucl-th/0601086}
  \BibitemShut {NoStop}%
\bibitem [{\citenamefont {{Page}}\ \emph {et~al.}(2013)\citenamefont {{Page}},
  \citenamefont {{Lattimer}}, \citenamefont {{Prakash}},\ and\ \citenamefont
  {{Steiner}}}]{Page2013}%
  \BibitemOpen
  \bibfield  {author} {\bibinfo {author} {\bibfnamefont {D.}~\bibnamefont
  {{Page}}}, \bibinfo {author} {\bibfnamefont {J.~M.}\ \bibnamefont
  {{Lattimer}}}, \bibinfo {author} {\bibfnamefont {M.}~\bibnamefont
  {{Prakash}}}, \ and\ \bibinfo {author} {\bibfnamefont {A.~W.}\ \bibnamefont
  {{Steiner}}},\ }in\ \href@noop {} {\emph {\bibinfo {booktitle} {Novel
  Superfluids}}},\ \bibinfo {series and number} {International Series of
  Monographs on Physics},\ \bibinfo {editor} {edited by\ \bibinfo {editor}
  {\bibfnamefont {K.~H.}\ \bibnamefont {{Bennemann}}}\ and\ \bibinfo {editor}
  {\bibfnamefont {J.~B.}\ \bibnamefont {{Ketterson}}}}\ (\bibinfo  {publisher}
  {Oxford University Press},\ \bibinfo {address} {Oxford, UK},\ \bibinfo {year}
  {2013})\ p.\ \bibinfo {pages} {505}\BibitemShut {NoStop}%
\bibitem [{\citenamefont {{Sedrakian}}\ and\ \citenamefont
  {{Clark}}(2018)}]{2018arXiv180200017S}%
  \BibitemOpen
  \bibfield  {author} {\bibinfo {author} {\bibfnamefont {A.}~\bibnamefont
  {{Sedrakian}}}\ and\ \bibinfo {author} {\bibfnamefont {J.~W.}\ \bibnamefont
  {{Clark}}},\ }\href@noop {} {\bibfield  {journal} {\bibinfo  {journal} {arXiv:1802.00017,
  }\ } }\ \Eprint
  {http://arxiv.org/abs/1802.00017} {arXiv:1802.00017 [nucl-th]} \BibitemShut
  {NoStop}%
\bibitem [{\citenamefont {{Blaschke}}\ \emph {et~al.}(2004)\citenamefont
  {{Blaschke}}, \citenamefont {{Grigorian}},\ and\ \citenamefont
  {{Voskresensky}}}]{2004A&A...424..979B}%
  \BibitemOpen
  \bibfield  {author} {\bibinfo {author} {\bibfnamefont {D.}~\bibnamefont
  {{Blaschke}}}, \bibinfo {author} {\bibfnamefont {H.}~\bibnamefont
  {{Grigorian}}}, \ and\ \bibinfo {author} {\bibfnamefont {D.~N.}\ \bibnamefont
  {{Voskresensky}}},\ }\href {\doibase 10.1051/0004-6361:20040404} {\bibfield
  {journal} {\bibinfo  {journal} {\aap}\ }\textbf {\bibinfo {volume} {424}},\
  \bibinfo {pages} {979} (\bibinfo {year} {2004})},\ \Eprint
  {http://arxiv.org/abs/arXiv:astro-ph/0403170} {arXiv:astro-ph/0403170}
  \BibitemShut {NoStop}%
\bibitem [{\citenamefont {{Blaschke}}\ \emph {et~al.}(2012)\citenamefont
  {{Blaschke}}, \citenamefont {{Grigorian}}, \citenamefont {{Voskresensky}},\
  and\ \citenamefont {{Weber}}}]{2012PhRvC..85b2802B}%
  \BibitemOpen
  \bibfield  {author} {\bibinfo {author} {\bibfnamefont {D.}~\bibnamefont
  {{Blaschke}}}, \bibinfo {author} {\bibfnamefont {H.}~\bibnamefont
  {{Grigorian}}}, \bibinfo {author} {\bibfnamefont {D.~N.}\ \bibnamefont
  {{Voskresensky}}}, \ and\ \bibinfo {author} {\bibfnamefont {F.}~\bibnamefont
  {{Weber}}},\ }\href {\doibase 10.1103/PhysRevC.85.022802} {\bibfield
  {journal} {\bibinfo  {journal} {\prc}\ }\textbf {\bibinfo {volume} {85}},\
  \bibinfo {eid} {022802} (\bibinfo {year} {2012})},\ \Eprint
  {http://arxiv.org/abs/1108.4125} {arXiv:1108.4125 [nucl-th]} \BibitemShut
  {NoStop}%
\bibitem [{\citenamefont {{Beloin}}\ \emph {et~al.}(2018)\citenamefont
  {{Beloin}}, \citenamefont {{Han}}, \citenamefont {{Steiner}},\ and\
  \citenamefont {{Page}}}]{2018PhRvC..97a5804B}%
  \BibitemOpen
  \bibfield  {author} {\bibinfo {author} {\bibfnamefont {S.}~\bibnamefont
  {{Beloin}}}, \bibinfo {author} {\bibfnamefont {S.}~\bibnamefont {{Han}}},
  \bibinfo {author} {\bibfnamefont {A.~W.}\ \bibnamefont {{Steiner}}}, \ and\
  \bibinfo {author} {\bibfnamefont {D.}~\bibnamefont {{Page}}},\ }\href
  {\doibase 10.1103/PhysRevC.97.015804} {\bibfield  {journal} {\bibinfo
  {journal} {\prc}\ }\textbf {\bibinfo {volume} {97}},\ \bibinfo {eid} {015804}
  (\bibinfo {year} {2018})}\BibitemShut {NoStop}%
\end{thebibliography}

%

\end{document}